\documentclass[fleqn,usenatbib]{mnras}

\usepackage{newtxtext,newtxmath}
\usepackage[T1]{fontenc}

\DeclareRobustCommand{\VAN}[3]{#2}
\let\VANthebibliography\thebibliography
\def\thebibliography{\DeclareRobustCommand{\VAN}[3]{##3}\VANthebibliography}

\usepackage{graphicx}	% Including figure files
\usepackage{amsmath}	% Advanced maths commands
\usepackage{orcidlink}
\usepackage{soul}
\newcommand*{\mailto}[1]{\href{mailto:#1}{#1}}
\usepackage[defaultcolor=red]{changes}

\title[Predictions for ultra-faint dwarfs using GALACTICUS]{A comprehensive model for the formation and evolution of the faintest Milky Way dwarf satellites}

\author[N. Ahvazi et al.]{Niusha Ahvazi,$^{\orcidlink{0009-0002-1233-2013}1,2}$\thanks{E-mail: \mailto{niusha.ahvazi@email.ucr.edu} / \mailto{nahvazi@carnegiescience.edu}}
Andrew Benson,$^{\orcidlink{0000-0001-5501-6008}2}$
Laura V. Sales,$^{\orcidlink{0000-0002-3790-720X}1}$
Ethan O.~Nadler,$^{\orcidlink{0000-0002-1182-3825}2,3}$
Sachi Weerasooriya,$^{\orcidlink{0000-0001-9485-6536}2}$\and
Xiaolong Du,$^{\orcidlink{0000-0003-0728-2533}4,2}$
and Mia Sauda Bovill$^{\orcidlink{0000-0003-4037-5360}5}$ 
\\\\
$^{1}$Department of Physics and Astronomy, University of California, Riverside, CA 92521, USA\\
$^{2}$Carnegie Observatories, 813 Santa Barbara Street, Pasadena, CA 91101, USA\\
$^{3}$Department of Physics \& Astronomy, University of Southern California, Los Angeles, CA, 90007, USA\\
$^{4}$Department of Physics and Astronomy, University of California, Los Angeles, CA 90095, USA \\
$^{5}$Department of Astronomy, University of Maryland, College Park, MD 20742, USA
}

\date{Accepted XXX. Received YYY; in original form ZZZ}

\pubyear{2023}

\begin{document}
\label{firstpage}
\pagerange{\pageref{firstpage}--\pageref{lastpage}}
\maketitle

\begin{abstract}
In this study, we modify the semi-analytic model {\sc Galacticus} in order to accurately reproduce the observed properties of dwarf galaxies in the Milky Way. We find that reproducing observational determinations of the halo occupation fraction and mass-metallicity relation for dwarf galaxies requires us to include H$_2$ cooling, an updated UV background radiation model, and to introduce a model for the metal content of the intergalactic medium. By fine-tuning various model parameters and incorporating empirical constraints, we have tailored the model to match the statistical properties of Milky Way dwarf galaxies, such as their luminosity function and size--mass relation. We have validated our modified semi-analytic framework by undertaking a comparative analysis of the resulting galaxy-halo connection. We predict a total of  $300 ^{+75} _{-99}$ satellites with an absolute $V$-band magnitude (M$_{V}$) less than 0 within $300$ kpc from our Milky Way-analogs. The fraction of subhalos that host a galaxy at least this bright drops to $50\%$ by a halo peak mass of $\sim 8.9 \times 10^{7}$ M$_{\odot}$, consistent with the occupation fraction inferred from the latest observations of Milky Way satellite population. 
\end{abstract}


%%%%%%%%%%%%%%%%%%%%%%%%%%%%%%%%%%%%%%%%%%%%%%%%%%

%%%%%%%%%%%%%%%%% BODY OF PAPER %%%%%%%%%%%%%%%%%%

\section{Introduction}

Dwarf galaxies, characterized by their low masses, hold a prominent position in astrophysical research due to their intriguing properties and profound implications for our understanding of galaxy formation and evolution \citep{Simon2019}. From a theoretical perspective, these faint stellar systems offer valuable insights into fundamental aspects of galaxy formation models and cosmological paradigms \citep{doi:10.1146/annurev-astro-091916-055313, Sales2022}. One key reason for the significant interest in dwarf galaxies is their low mass and shallow gravitational potential wells, which makes them ideal laboratories for testing various feedback mechanisms. Feedback processes, such as stellar winds and supernovae play a crucial role in regulating star formation and shaping the properties of galaxies \citep{Bower2012, Zolotov2012, Puchwein2013, Madau_2014, Chan2015, Read2016, Tollet2016, Fitts2017}. Dwarf galaxies, with their shallower gravitational potentials provide an excellent testing ground to investigate the interplay between these feedback processes and the surrounding circumgalactic medium \citep{Lu_2017,2018ApJ...867..142C}. Their formation predates that of more massive galaxies, allowing us a glimpse of the conditions and processes that prevailed during the early stages of the Universe. For example, the low metallicity exhibited by dwarf galaxies presents an opportunity to probe the mechanisms responsible for chemical enrichment in the early Universe \citep{2009ApJ...693.1859B,2011ApJ...741...17B,2015MNRAS.453.1305W}.  By studying these ancient systems, we gain valuable insights into the hierarchical assembly of galaxies and the mechanisms responsible for their subsequent evolution.

In addition, the study of dwarf galaxies contributes to our understanding of the nature of dark matter (DM; e.g. \citealt{Maccio2019, Nadler2021, Newton2021, Dekker2022}). As the most numerous galaxy population in the Universe \citep{Ferguson1994}, their abundance and distribution provide essential constraints for cosmological models, particularly those based on cold dark matter (CDM). By investigating the properties and spatial distribution of dwarf galaxies, we can test the predictions of the CDM model and explore alternative models that may better explain their observed characteristics.

In tandem with theoretical interest, there has been a remarkable growth in the observational landscape of dwarf galaxies over the past two decades---from surveys\footnote{We refer the reader to \protect\cite{Crnojevi2021} for examples of discovered dwarfs in each survey.} including the Sloan Digital Sky Survey (SDSS; \citealt{Ahumada2020, Abdurro2022, Almeida2023}), Dark Energy Survey (DES; \citealt{Bechtol2015, Drlica-Wagner2015}), The DECam Local Volume Exploration Survey (DELVE; \citealt{Drlica2022Survey}), Pan-STARRS (PS1; \citealt{Chambers2016}), ATLAS \citep{Shanks2015}, and Gaia \citep{GaiaCollaboration2016}. Advancements in survey capabilities and data analysis techniques have led to a significant increase in the number of known Milky Way (MW) dwarfs, enabling a detailed characterization of their properties. Relevant examples that target MW or MW-like environments in the Local Volume include \protect\cite{Geha2017, Mao2021, Carlsten2021, Nashimoto2022, Danieli2017, Bennet2020, Doliva-Dolinsky2023, Smercina2018}. These observations have provided crucial empirical constraints for theoretical models and paved the way for a deeper understanding of the formation and evolution of dwarf galaxies.

The motivation behind this paper is to construct a comprehensive, physical model that accurately reproduces the statistical properties of MW dwarf galaxies. Therefore, by developing this model, we can shed light on the underlying physics and unravel the intricate mechanisms that govern the formation and evolution of these galaxies. Furthermore, our motivation extends beyond the mere reproduction of observed statistical properties. We also seek to investigate how dwarf galaxies respond to changes in the nature of DM. To explore the impact of DM on dwarf galaxies, it is imperative to begin with a model that accurately represents the prevailing cosmological paradigm, specifically the CDM model. By establishing a reliable foundation based on CDM, we can examine how variations in the nature of dark matter affect the properties of dwarf galaxies (specifically, the self-interacting dark mater model, Ahvazi et al. in prep.). This endeavor enables us to probe the sensitivity of dwarf galaxies to different DM scenarios, providing crucial insights into the nature and fundamental properties of dark matter itself.

In this study, we adopt a systematic approach by modifying the existing Semi-Analytic Model (SAM) known as ``{\sc Galacticus}'' \citep{Benson2012Galacticus} to accurately reproduce the observed properties of dwarf galaxies in the MW. The SAM framework serves as a powerful tool for establishing the connection between the formation and evolution of galaxies and the underlying dark matter halos in which they reside. One notable advantage of the SAM approach is its computational efficiency, enabling us to explore numerous realizations and, in the future, investigate different dark matter physics rapidly \citep{Benson2012Galacticus, Benson2013}. By employing the SAMs, we can effectively resolve dwarfs and ultra-faints within much more massive systems, including clusters, which are typically beyond the reach of hydrodynamic simulations \citep{Pillepich2019, Nelson2019, Tremmel2019}. It should be noted, however, that for MW-like systems, the latest generation of zoom-in hydrodynamical simulations are achieving resolutions sufficient for resolving ultra-faint dwarf galaxies \citep{Buck2020, Applebaum2021, Grand2021, Joshi2024}. It is crucial to recognize that while hydrodynamical simulations, in principle, offer higher accuracy by relying on fewer assumptions, their computational demands are substantially larger than those of SAMs.

Our first objective is to tailor the SAM to match the statistical properties of MW dwarf galaxies, such as their luminosity function and metallicities, by carefully adjusting various model parameters and incorporating empirical constraints. In addition, we include models that we anticipate will play a pivotal role in the evolution of dwarf galaxies. Specifically, we incorporate H$_2$ cooling and consider the influence of Intergalactic Medium (IGM) metallicity, and UV background radiation. H$_2$ cooling is particularly significant in low-mass halos, as it affects the ability of gas to condense and form stars. Furthermore, the inclusion of IGM metallicity enables us to account for the metal enrichment of dwarf galaxies more accurately.

To assess the implications of our modifications and refinements, we undertake a comparative analysis of the resulting galaxy--halo connection. This step is crucial as it enables us to investigate the relationship between the observed properties of dwarf galaxies and the underlying dark matter halos. By comparing our results with prior estimates of this connection, we gain insights into the distribution of dark matter within dwarf galaxies and its impact on their observable characteristics. This comparison also serves as a validation of our modified SAM framework and allows us to assess the extent to which our model aligns with existing knowledge and understanding of the galaxy--halo connection in the context of MW dwarf galaxies. Moreover, we leverage our model to make predictions for the mass function of halos across a range of masses, encompassing ultra-faint satellites of Large Magellanic Cloud (LMC)-analogs, satellites of M31-analog systems, as well as dwarfs residing in group and cluster environments. 

This paper is organized as follows: In Section~\ref{methods}, we provide a detailed description of our methodology, outlining the modifications made to the existing {\sc Galacticus} model and the incorporation of key physical processes. In Section~\ref{results}, we present our comprehensive results and engage in a discussion of the galaxy--halo connection, in Section~\ref{Galaxy--halo connection}. we present our predictions for various quantities associated with dwarf galaxies, in Section~\ref{Dwarf population}, and we explore the mass functions of halos across different mass scales, in Section~\ref{Mass function predictions}. Finally, in Section ~\ref{Conclusions}, we summarize our significant findings and draw conclusions based on the analysis conducted in this study. 

\section{Methods}\label{methods}

We use the {\sc Galacticus} semi-analytical model (SAM) for galaxy formation and evolution as introduced by \cite{Benson2012Galacticus}.\footnote{The specific version used in this work is publicly available at \href{https://github.com/galacticusorg/galacticus/commit/b60e818869ea0bad7e2fcc2b9320cabbe02cf550}{https://github.com/galacticusorg/galacticus}.} Similar to other SAMs---including the Santa Cruz SAM \citep{Somerville1999}, GALFORM \citep{Cole2000}, SAG \citep{Cora2006}, MORGANA \citep{Monaco2007}, L-Galaxies \citep{Henriques2015}, SAGE \citep{Croton2016}, and SHARK \citep{2018MNRAS.481.3573L}---{\sc Galacticus} parameterizes the astrophysical processes that govern galaxy formation and evolution and uses a set of differential equations to model and solve galactic evolution over time. It builds dark matter halo merger trees by employing a modified extended Press-Schechter formalism \citep{Parkinson2007, 2017MNRAS.467.3454B} and then simulates the evolution of galaxy populations within this merging hierarchy of halos. At the end of this evolution process, {\sc Galacticus} provides a comprehensive set of properties for the galaxies, including stellar mass, size, metallicity, morphology, star formation history, and photometric luminosities derived using simple stellar population spectra from the FSPS model\footnote{\href{https://github.com/cconroy20/fsps/releases/tag/v3.2}{https://github.com/cconroy20/fsps/releases/tag/v3.2}} \citep{Conroy2009}. 

The baryonic physics of the Galacticus model has been constrained by adjusting parameters to match a variety of observational data on massive galaxies (typically $L_*$ and brighter systems) as described in \protect\citeauthor{2018MNRAS.474.5206K}~(\protect\citeyear{2018MNRAS.474.5206K}; Section~2.2), which also summarizes the key baryonic physics in Galacticus. Parameter tuning was performed by manually searching the model parameter space to seek models that closely match observations including the $z=0$ stellar mass function of galaxies, $z=0$ luminosity functions, the local Tully-Fisher relation, distributions of galaxy colors and sizes, the black hole--bulge mass relation, and the star formation history of the universe. \protect\cite{2018MNRAS.474.5206K} also presents a number of comparisons between the predictions of Galacticus and observations for massive galaxies. These comparisons show that Galacticus performs well in matching observational estimates of the distribution of star formation rates in galaxies, the cosmic star formation history, the distribution of black hole masses, the stellar mass--halo mass relation, and measures of galaxy clustering. However, in other comparisons (e.g. galaxy cold gas content and metallicity), Galacticus fares less well against the observational constraints.

{\sc Galacticus} is designed to be highly modular, and offers the flexibility to incorporate various models for the complex processes involved in galaxy formation and evolution. Starting from the model presented in \protect\cite{2018MNRAS.474.5206K}, in this work, we utilize a model similar to that recently proposed by \cite{Weerasooriya2022}, but with some differences. In contrast to \cite{Weerasooriya2022}, who utilized merger trees extracted from N-body simulations and ran {\sc Galacticus} on those trees, we employ the merger tree building algorithm of \cite{Cole2000}, which is based on the extended Press-Schechter (EPS) formalism, with the modifier function proposed by \cite{Parkinson2007}---recalibrated to improve the match to high-resolution zoom-in simulations of Milky Way mass halos (Sarnaaik et al., in prep.). We combine this with a comprehensive subhalo evolution model in {\sc Galacticus}. The rationale behind this choice is our aim to generate a large number of realizations of MW-analogs, while fully-resolving halos hosting the lowest mass galaxies, allowing us to investigate the effects of baryons on galaxy properties. Additionally, in upcoming papers, we plan to explore the implications of different dark matter models and the presence of an LMC-analog.

Given our aim of comprehensively studying the entire MW dwarf population (down to ultra-faints) in this paper, the effects of resolution become particularly important. A key consideration is the impact of resolution on the results obtained by \cite{Weerasooriya2022}, as they discussed in Section 3.3.1 of their paper---their merger trees resolved $10^7\mathrm{M}_\odot$ halos with just 100 particles. The resolution of N-body simulations can limit the ability to predict sizes for low-mass dwarfs accurately. 

In addition to the resolution difference, other distinctions between these two approaches include the treatment of the effect of reionization and the suppression of gas accretion into low-mass halos. While \cite{Weerasooriya2022} utilized a simple model involving sharp cuts in virial velocity to mimic these effects, we opt for a more realistic model in our work (see Appendix~\ref{app:model}). Moreover, we adopt different cooling rates, feedback mechanisms, a reionization model, and accretion mode, along with specific angular momentum prescriptions, as explained in detail in Appendix~\ref{app:model}. Despite employing this more realistic model, we maintain the same level of agreement with observational results and predictions inferred from observational data, ensuring the robustness and reliability of our findings. For a brief comparison with other SAM approaches, the reader is referred to Appendix~\ref{app:otherSAMs}.

In our model, we employ a comprehensive treatment for the orbital evolution of subhalos, incorporating essential nonlinear dynamical processes, including dynamical friction, tidal stripping, and tidal heating. This model was first implemented in {\sc Galacticus} by \citet[][the reader is referred to \citealt{Yang2020} for a full explanation and an initial calibration of the model]{Pullen2014}. Subsequently, the tidal heating model was improved by \cite{Benson2022} to include second-order terms in the impulse approximation which is shown to more accurately follow the tidal tracks measured in high-resolution N-body simulations. For a comprehensive and detailed account of the subhalo orbital evolution within our model, please refer to Appendix~\ref{app:model_A2}. In addition to providing a more detailed treatment of the evolution of subhalo density profiles, the primary advantage of this treatment of subhalo orbits for the present work is that it provides orbital radii for all subhalos, allowing us to select satellite galaxies based on their distance from the MW. Furthermore, the central galaxy in our model is evolved self-consistently, following the same baryonic physics (e.g. star formation, feedback, etc) as described for the evolution of subhalos. Importantly, the gravitational potential of the MW is included at all times when modeling our subhalo orbital evolution, providing a more accurate representation of the gravitational interactions between the central galaxy and its satellite subhalos. 

In this study, we track the evolution of 100 MW-analogs or host halos with $z=0$ masses ranging from $7 \times 10^{11}$ to $1.9 \times 10^{12}$ $\rm \mathrm{M}_\odot$ \citep{Wang2020,Callingham2019}, and resolving progenitor halos to masses of $10^{7}$ $\rm \mathrm{M}_\odot$---sufficient to fully resolve the formation of ultra-faint  dwarf galaxies similar to those observed in the vicinity of the MW as we will demonstrate below. To calibrate and test our models of MW-analogs and their subhalo population, we use observational data from Local Group dwarf galaxies, including all Milky Way dwarf galaxies from the DES+PS1 surveys \citep{Drlica-Wagner2020} and the updated \cite{McConnachie2012} compilation, along with ultra-faint dwarf population from \cite[see references therein]{Simon2019}, and few extra objects such as Pegasus IV \citep{Cerny2023}, Indus I \citep{Koposov2015a}, Antlia II \citep{Torrealba2019}, and Centaurus I \citep{Mau2020}.

A primary advantage of using a semi-analytic approach is its computational efficiency, which enables rapid exploration of parameter space and model space. This allows for the study of the effects of various models on the evolution of halos and galaxies. In this paper, we focus on examining the effects of the redshift evolution of the IGM metallicity, the effect of different models of the cosmic UV background radiation, and the contribution of molecular hydrogen, H$_2$, to the cooling function of CGM gas. We present the implementation details of these models in sections \ref{sec:IGMmetallicity}, \ref{sec:UVradiation}, and \ref{sec:h2Cooling} respectively.

\subsection{IGM metallicity} \label{sec:IGMmetallicity}

The presence of metals in the IGM has been confirmed through observations, indicating their existence at significant levels during the redshifts corresponding to dwarf galaxy formation \citep{Madau2014, Aguirre2008, Simcoe2004, Schaye2003}. In addition, studies of dwarf galaxies have revealed a noticeable plateau in the mass-metallicity relation at lower masses \citep{Simon2019}. Our feedback model, which follows a power law dependence on the gravitational potential of galaxies (and so, for dwarf galaxies, is close to a power law dependence on halo mass), does not inherently produce such a plateau in the mass-metallicity relation ---instead it results in an effective yield (and, therefore, a stellar metallicity) that decreases continuously toward lower halo masses \citep[see, for example][\S4.2.1 \& \S4.2.2]{Cole2000}. Motivated by these facts, we propose that the metallicity of the IGM might play a crucial role in shaping the mass-metallicity relation of galaxies, and may potentially explain the observed plateau. In light of this hypothesis, we introduce a simple model that incorporates the metallicity of the IGM, aiming to elucidate the underlying mechanisms that govern the observed plateau. By considering the impact of IGM metallicity on the evolution of dwarf galaxies, we can gain valuable insights into the interplay between the metal enrichment of the IGM and the metallicity of inflowing material. It is important to note that the detailed mechanisms responsible for enriching the IGM with metals, including the propagation and mixing of outflows, remain subjects of ongoing theoretical investigation (\citealt{2020MNRAS.494.3971M,2017MNRAS.468.4170M,2020ApJ...895...43S}; see \citealt{2017ARA&A..55..389T} for a comprehensive review), and we do not attempt to model them here.

Therefore, this study uses a simple polynomial model to describe how the IGM metallicity evolves as a function of redshift. Specifically, we assume that the metallicity is given by
\begin{equation}\label{eq:IGMmetal}
    \log_{10} (Z_\mathrm{IGM}/\mathrm{Z}_\odot) = A + B \log_{10} (1+z),
\end{equation}
where $Z_\mathrm{IGM}$ represents the metallicity of the IGM and $z$ is redshift. This model incorporates two free parameters, $A$ and $B$ that are calibrated to match current observations of the mass-metallicity relation of dwarf galaxies and to satisfy inferences on $Z_\mathrm{IGM}$ from observations of the $\rm Ly\alpha$ forest in the spectra of distant quasars.

\subsection{UV background radiation} \label{sec:UVradiation}

The cosmic background of UV radiation plays a key role in the evolution of molecular hydrogen in low-mass halos through the process of photodissociation (see \S\ref{sec:h2Cooling} below). A key factor for our work is the redshift at which reionization of the IGM occurs. After reionization the UV background radiation is able to increase in intensity substantially (as the IGM becomes transparent at these wavelengths), resulting in greatly enhanced photodissociation of molecular hydrogen.

In this work we make use of two models of the cosmic background radiation---with significantly different reionization redshifts---to allow us to explore how our results depend on this choice. 

The first model we consider is that of \defcitealias{2012ApJ...746..125H}{HM12}\cite{2012ApJ...746..125H} (HM12 hereafter). This model includes a ``minimal reionization model'' which was shown to produce an optical depth to reionization of $\tau_\mathrm{es} = 0.084$ in good agreement with the (current at the time of publication of \citetalias{2012ApJ...746..125H}) WMAP 7-year results of $\tau_\mathrm{es}=0.088\pm0.015$ \citep{2011ApJS..192...14J}, and a reionization redshift (the epoch at which the volume filling fraction of HII reaches 50\%) of $z\approx 10$.

The second model that we use is that of \defcitealias{2020MNRAS.493.1614F}{FG20}\cite{2020MNRAS.493.1614F} (FG20 hereafter) which is calibrated to more recent data (a complete discussion, and comparison to earlier works, is given in the paper). Importantly for our work, the \cite{2020MNRAS.493.1614F} model produces an optical depth to reionization of $\tau_\mathrm{es} = 0.054$, matched to that measured by the Planck 2018 analysis, $\tau_\mathrm{es} = 0.054\pm0.007$ \citep{2020A&A...641A...6P}, and therefore a lower reionization redshift of $z=7.8$.

We consider \cite{2020MNRAS.493.1614F} to be the preferred model for the cosmic background radiation (as it is calibrated to more accurate measures of the optical depth to reionization), but explore the \cite{2012ApJ...746..125H} model also to investigate how the redshift of reionization affects our results.

In both cases, the spectral radiance of the cosmic background radiation is computed by interpolating in tables (as a function of wavelength and redshift) derived from these two models.

\subsection{Molecular hydrogen cooling}\label{sec:h2Cooling}

In halos with virial temperatures below the atomic cooling cut off (at around $10^4$~K), the primary coolant for gas in the circumgalactic medium of high-redshift halos is molecular hydrogen (H$_2$; e.g. \citealt{1995PhDT........85A,1997ApJ...474....1T}). Even with our added pre-enrichment in the IGM metallicity (see section~\ref{sec:IGMmetallicity}), the metallicity of the cooling case remains sufficiently low that the metal line cooling is not substantially enhanced, while H$_2$ becomes sufficiently abundant at T $<$ 10$^4$ K.\footnote{While this is true for the model presented in this work, we caution readers that the outcomes may be sensitive to the underlying assumptions in computing metal cooling and H$_2$ formation/destruction in the presence of a radiation field. For instance, a comparison of our cooling efficiencies with \protect\cite{Bialy2019} reveals overall agreement at the typical densities of our halos, although they emphasize the impact of the surrounding radiation field, particularly the susceptibility of H$_2$ to destruction by the far-UV radiation (see their Fig. 7, top panels), and the strong density dependence in the contribution of H$_2$ to cooling. The efficiency of H$_2$ cooling in small, early-forming halos, considering photodissociation through Lyman-Werner radiation in the presence of H$_2$ self-shielding, remains a debated topic in the literature (see Section 4.3.2 of the review by \citealt{Klessen2023}, and references therein). In general, the actual efficiency and relevance of H$_2$ cooling in small, early-forming halos are subjects of ongoing debate.} The timescales of the reactions which form and destroy H$_2$ can be long compared to halo assembly timescales, meaning that equilibrium abundances can not be assumed. Therefore, we must solve the rate equations for the production and destruction of H$_2$ in each halo. This is straightforward as these can simply be added as additional equations passed to {\sc Galacticus}' differential equation solver engine which then integrates them forward in time with adaptive timesteps chosen to achieve a suitable accuracy.

We use the network of chemical reactions described in \cite{1997NewA....2..181A} to track the abundance of H$_2$---in particular we follow their recommendation for a ``fast'' network by assuming that H$^-$ is always present at its equilibrium abundance and ignoring various slow reactions. Therefore, in the circumgalactic medium (CGM) of each halo we track the abundances of H, H$^+$, H$_2$, and e$^-$, and include the following set of reactions:
\begin{itemize}
\item $\hbox{H} + \hbox{e}^- \rightarrow \hbox{H}^+ + 2 \hbox{e}^-$;
\item $\hbox{H}^+ + \hbox{e}^- \rightarrow \hbox{H} + \gamma$;
\item $\hbox{H} + \hbox{H}^- \rightarrow \hbox{H}_2 + \hbox{e}^-$;
\item $\hbox{H}_2 + \hbox{e}^- \rightarrow 2\hbox{H} + \hbox{e}^-$;
\item $\hbox{H}^- + \gamma \rightarrow \hbox{H} + \hbox{e}^-$;
\item $\hbox{H}_2 + \gamma \rightarrow H_2^* \rightarrow 2\hbox{H}$;
\item $\hbox{H}_2 + \gamma \rightarrow 2\hbox{H}$; and
\item $\hbox{H} + \gamma \rightarrow \hbox{H}^+ +\hbox{e}^-$,
\end{itemize}
utilizing the rate coefficients and cross sections given by \cite{1997NewA....2..181A} in each case. The temperature of the CGM is assumed to be equal to the virial temperature of the halo for the purposes of computing rate coefficients (and for the purposes of computing cooling functions---see below).

In computing the evolution of the abundances we assume a uniform density CGM, in which the current CGM mass is contained within a sphere of radius $r_\mathrm{CGM}$ which we take to be the virial radius for halos, and the ram pressure radius for subhalos\footnote{{\sc Galacticus} implements the ram pressure stripping model of \protect\cite{2008MNRAS.389.1619F} as described in \protect\cite{2015ApJ...799..171B}. As the mass of the CGM in a subhalo is reduced due to the effects of ram pressure stripping from the CGM of its host halo, we assume that this mass is removed in spherical shells from the subhalo CGM, starting at the outer edge, $r_\mathrm{CGM}$. In this way, the outer edge, $r_\mathrm{CGM}$, decreases over time as ram pressure stripping proceeds.}. However, we account for the fact that the CGM will be denser in the inner regions of the halo via a clumping factor, $f_\mathrm{c}$, which multiplies the rates of the first three reactions (i.e. those involving two CGM particles). The clumping factor is computed as
\begin{equation}
    f_\mathrm{c} = \frac{\langle \rho_\mathrm{CGM}^2 \rangle}{\langle \rho_\mathrm{CGM} \rangle^2} = \frac{4 \pi r_\mathrm{CGM}^3}{3 M_\mathrm{CGM}^2} \int_0^{r_\mathrm{CGM}} 4 \pi r^2 \rho_\mathrm{CGM}^2(r) \mathrm{d} r,
\end{equation}
where $\langle\rangle$ indicates a volume average, and $\rho_\mathrm{CGM}$ is the density of the CGM, which we model as a $\beta$-profile with core radius equal to 30\% of the virial radius.

In computing the rate for the reaction $\hbox{H}_2 + \gamma \rightarrow H_2^* \rightarrow 2\hbox{H}$ we account for self-shielding of the radiation following the model of \citeauthor{2012MNRAS.426.1159S}~(\citeyear{2012MNRAS.426.1159S}; their eqn.~11), estimating the H$_2$ column density at $N_\mathrm{H_2} \approx n_\mathrm{H_2} r_\mathrm{CGM}$, where $n_\mathrm{H_2}$ is the density of H$_2$ in the CGM.

Solving the network of reactions to compute the H$_2$ abundance can be computationally demanding. In particular, in higher mass halos (at higher temperatures) the timescales for the reactions controlling the ionization state of atomic hydrogen can become very short, requiring a large number of small timesteps to solve. However, in such cases the ionization fraction of atomic hydrogen rapidly approaches its equilibrium value and, furthermore, the abundance of H$_2$ is typically very low in such halos as it is destroyed by collisions at high temperatures, meaning that it makes little contribution to the cooling function. Therefore, we choose to switch over to an equilibrium calculation when
\begin{equation}
 \tau_\mathrm{H} < f_\mathrm{dyn} \tau_\mathrm{dyn},
\end{equation}
where $f_\mathrm{dyn}$ is a parameter, $\tau_\mathrm{dyn}$ is the dynamical time in the halo\footnote{Dynamical time here is defined as $\tau_\mathrm{dyn}$ = $\sqrt{r_\mathrm{v}^3/\mathrm{G} M_\mathrm{v}}$, where r$_\mathrm{v}$ and M$_\mathrm{v}$ are virial radius and virial mass of the halo, respectively.}, and
\begin{equation}
  \tau_\mathrm{H} = \mathrm{min}\left(\tau_\alpha,\tau_\beta,\tau_\Gamma\right),
\end{equation}
where $\tau_\alpha=1/\alpha n$, $\tau_\beta=1/\beta n$,$\tau_\Gamma=1/\Gamma$, $n$ is the number density of hydrogen, and $\alpha$, $\beta$, and $\Gamma$ are the collisional ionization, radiative recombination, and photoionization rate coefficients for hydrogen respectively.

If the system is judged to be in equilibrium then the neutral fraction of hydrogen is computed as:
\begin{equation}
 x_\mathrm{H} = \frac{ \tau_\Gamma^{-1} + \tau_\alpha^{-1} +2 \tau_\beta^{-1} - \sqrt{ \tau_\Gamma^{-2} +2 \tau_\alpha^{-1} \tau_\Gamma^{-1}  + \tau_\alpha^{-2}  + 4 \tau_\beta^{-1} \tau_\Gamma^{-1} } }{ 2 \tau_\alpha^{-1} + \tau_\beta^{-1} }.
\end{equation}
The abundances of H, H$^+$, and e$^-$ are then fixed according to this fraction, and reaction rates for them are set to zero. The reactions controlling the formation/destruction of H$_2$ are still followed as normal, by directly solving the relevant differential equations (but now using the equilibrium abundances for H, H$^+$, and e$^-$).

We use a value of $f_\mathrm{dyn} = 10^{-3}$, such that this equilibrium approximation is only used when the timescale controlling the ionization state of atomic hydrogen is less than 0.1\% of the halo dynamical time. We have checked that the resulting evolution of the H$_2$ abundance agrees closely with that obtained using a fully non-equilibrium calculation (but is orders of magnitude faster).

Given the abundance of H$_2$ we then compute its contribution to the cooling function, $\Lambda(T)$, following the approach of \cite{1998A&A...335..403G} using the fitting functions given in that work.

\section{Results and Discussion}\label{results}

While our semi-analytic model is relatively fast to run, conducting a full likelihood analysis using an approach such as MCMC becomes computationally infeasible due to the large number of parameters involved and the resulting need to make tens of thousands of evaluations of the model. Therefore, we pursued an alternative approach by manually fine-tuning the parameters to accurately replicate the properties of higher-mass galaxies, including the luminosity functions and the mass-metallicity relation. Given that our model already demonstrated reasonably close agreement with higher-mass galaxies, minor adjustments were sufficient to capture the behavior of lower-mass regimes.

However, for the incorporation of the novel aspect of IGM metallicity, we elected to employ a likelihood analysis utilizing a coarse grid search and full likelihood calculations. This decision was motivated by computational tractability since this new aspect introduced only two parameters and was expected to primarily impact the metallicities of ultra-faint dwarfs, with minimal effects on the more massive systems already calibrated. Employing this methodology allowed us to determine the optimal values for the coefficients A and B (as introduced in equation~\ref{eq:IGMmetal}), yielding A = -1.3 and B = -1.9\footnote{The decision to use a coarse grid was primarily due to the computational expense associated with more extensive analyses, such as MCMC, which would be necessary for a comprehensive exploration of all free parameters across all models in this SAM. Given the computational limitations, we focused on finding the optimum values for the free parameters in the IGM metallicity model. However, it is important to acknowledge that the coarse grid search resulted in insufficient information to calculate a meaningful theoretical uncertainty for these parameters. Despite this limitation, we have ensured that the optimization process has good coverage of the available parameter space to the best extent possible under the computational constraints.}.

%%%%%%%%%%%%%%%%
\begin{figure}
    \includegraphics[width=\columnwidth]{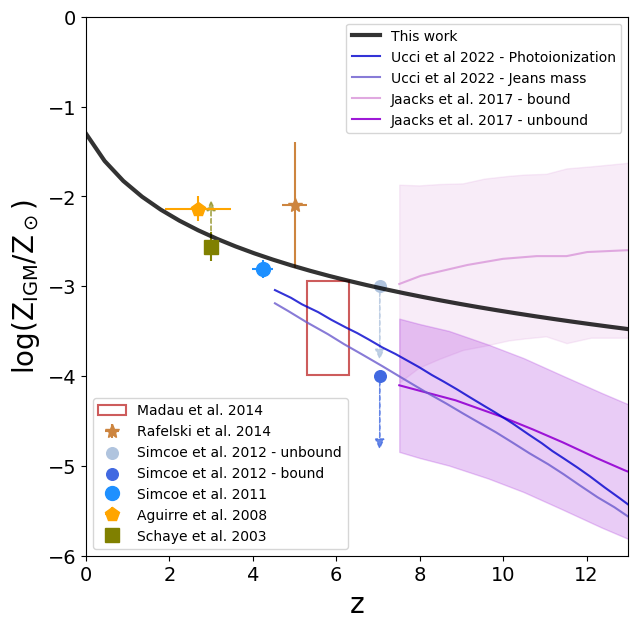}
    \caption{Evolution of IGM metallicity as a function of redshift. The black line represents the predicted evolution based on our model. Observational results are depicted by markers of different colors. The green square corresponds to the average [C/H] measurements reported by \protect\cite{Schaye2003}. The orange pentagon represents the metallicity of the IGM as probed by O VI absorption in the Ly$\alpha$ forest reported by \protect\cite{Aguirre2008}. The blue circles represent results by \protect\cite{Simcoe2011} and \protect\cite{Simcoe2012}. Additionally, the brown star marks measurements by \protect\cite{Rafelski2014} and, the red rectangle shows the carbon metallicity in the IGM as calculated by \protect\cite{Madau2014} based on observations from \protect\cite{Simcoe2011} and \protect\cite{Becker2011}. We also show results from cosmological hydrodynamical simulations. The simulations of \protect\citeauthor{Jaacks2018}~(\protect\citeyear{Jaacks2018}, which focus on Pop III modeling) are shown by pink and purple lines, while those of \protect\citeauthor{Ucci2023}~(\protect\citeyear[][which focus on modeling of reionization]{Ucci2023}) are shown by blue and violet lines.}
    \label{fig:IGMmetal}
\end{figure}
%%%%%%%%%%%%%%%%

Fig.~\ref{fig:IGMmetal} visually presents the variation of IGM metallicity with redshift as predicted by our model, represented by the black line. Additionally, we compared our model predictions with observations of IGM metallicity at higher redshifts. The average [C/H] measurements reported by \cite{Schaye2003} at redshift z = 3 yielded a value of -2.56, considering all their samples at this specific redshift, while not accounting for the effect of overdensity. However, focusing solely on quasars (rather than accounting for both galaxies and quasars in their sample) for determining the spectral shape of the metagalactic UV/X-ray background radiation resulted in measurements showing 0.5 dex higher values. Furthermore, \cite{Aguirre2008} examined the IGM metallicity probed by  O VI absorption in the Ly$\alpha$ forest for the redshift range 1.9 $<$ z $<$ 3.6 (represented by the orange marker). Observations by \protect\cite{Rafelski2014} at $4.7 < z < 5.3$ revealed metallicities ranging from [$-1.4$, $-2.8$]. The study by \cite{Madau2014} estimated carbon metallicity by probing C IV and C II absorption measurements from \cite{Simcoe2011} and \cite{Becker2011}, respectively, over the redshifts $5.3 - 6.4$. \cite{Madau2014} calculated the carbon metallicity assuming a range of [0.1, 1] for the ratio of singly or triply ionized carbon over this redshift range (depicted as the red rectangle on the plot). \protect\cite{Simcoe2011} explored IGM metallicity through C IV absorption in the redshift range $4 - 4.5$, while \protect\cite{Simcoe2012} reported chemical abundances of < 1/10,000 Solar if the gas is in a gravitationally bound proto-galaxy or <1/1,000 Solar if it is diffuse and unbound in a quasar spectrum at $z=7.04$, suggesting that gravitationally bound systems could be viable sites for the production of Pop III stars.

Turning to cosmological hydrodynamical simulations, \protect\cite{Jaacks2018} utilized the hydrodynamic and N-body code GIZMO coupled with their sub-grid Pop III model to study the baseline metal enrichment from Pop III star formation at $z > 7$ (results are shown in the figure by pink and purple lines corresponding to bound and unbound systems). Independently, the study by \protect\cite{Ucci2023} discusses the metal enrichment of the IGM at $z > 4.5$ through using a detailed physical model of galaxy chemical enrichment embedded into the {\sc astraeus} framework, which couples galaxy formation and reionization in the first billion years. Through their radiative feedback models, they explored a range from a weak, time-delayed (their ``Photoionization model'') to a strong instantaneous reduction of gas in the galaxy (their ``Jeans mass model''), with predictions shown on Fig.~\ref{fig:IGMmetal} by blue and violet lines, respectively\footnote{It is essential to treat the IGM metallicity values from \protect\cite{Ucci2023} as a lower limit since their method assumes that ejected metals are homogeneously dispersed into the entire simulation box when calculating $Z_\mathrm{IGM}$.}.

While observations appear to narrow down the range of IGM metallicities at lower redshifts, aligning with the expectation of our best model as determined through the likelihood analysis, uncertainties in modeling the metallicity evolution of the universe at higher redshifts prevent precise predictions of the metal content of the IGM. Predictions from our model suggest higher values of IGM metallicity at higher redshifts (the time of formation of ultra-faint galaxies) compared to the examples shown here. Hydrodynamical simulations generally predict IGM metallicities at high redshifts that are lower than those adopted in this work (and which we find are necessary to produce the correct metallicities of ultra-faint dwarf galaxies). However, we note that the simulation of \protect\cite{Jaacks2018} predicts substantially higher metallicities in bound regions. Given that the ultra-faint dwarfs studied in this work are, by definition, forming in a biased environment (the region around the proto-Milky Way), we may expect that they therefore experience a higher metallicity than that of the volume-averaged IGM. As such, while our IGM metallicity model remains empirical and speculative, it is within the bounds of current theory given the environment of interest.

\subsection{Galaxy--halo connection}\label{Galaxy--halo connection}

In this section, we explore the galaxy-halo connection and its sensitivity to the incorporation of molecular hydrogen cooling and UV background radiation, as introduced in sections \ref{sec:UVradiation} and \ref{sec:h2Cooling}. By examining the impact of these key physical processes on our sample of MW analogs, we aim to gain deeper insights into the intricate interplay between gas cooling, radiation, and galaxy formation within the context of our simulated galaxy population, particularly the low-mass dwarf satellites of our own MW. 

\subsubsection{Occupation fraction} \label{sec: occupation fraction}

\begin{figure}
    \includegraphics[width=\columnwidth]{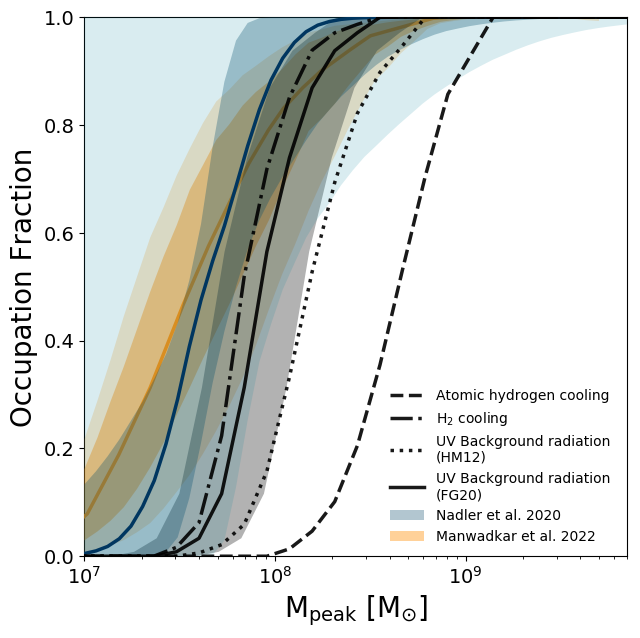}
    \caption{Occupation fraction as a function of the peak halo mass. The black curves, with different line styles, correspond to the predictions from our model incorporating various physical processes. Specifically, the dashed line corresponds to the model incorporating only atomic hydrogen cooling, while the dotted-dashed line represents the model incorporating both atomic and molecular hydrogen cooling (but no UV background radiation). The dotted line corresponds to the model including molecular hydrogen cooling and the UV background radiation prescription of \citetalias{2012ApJ...746..125H}. The black curve with a shaded gray region corresponds to the model including molecular hydrogen cooling and the UV background radiation prescription of \citetalias{2020MNRAS.493.1614F}. The gray shaded region indicates a 40\% uncertainty in estimating the peak masses from our simulation. Additionally, the blue curve, along with the dark and light shaded blue regions, corresponds to the predictions by \protect\citealt{Nadler2020}, while the orange curve, along with the dark and light shaded orange regions, corresponds to the predictions by \protect\citealt{Manwadkar2022}.}
    \label{fig:occupationFraction}
\end{figure}

The occupation fraction, a crucial measure of the galaxy-halo connection, is defined here to be the fraction of dark matter halos hosting a luminous galaxy with absolute $V$-band magnitudes less than 0, roughly equivalent to a stellar mass content greater than approximately $100 \mathrm{M}_\odot$. In Fig.~\ref{fig:occupationFraction}, we present the occupation fraction as a function of peak halo mass\footnote{In \textsc{Galacticus} halo masses are defined as overdense regions with a mean density equal to that predicted by the spherical collapse model for the adopted cosmology and redshift \citep{1980lssu.book.....P,1993MNRAS.262..627L,1996MNRAS.282..263E}.}. The dashed line represents the model incorporating only atomic hydrogen cooling, while the dotted-dashed line corresponds to the model including both atomic and molecular hydrogen cooling (but ignoring effects of the UV background radiation). A comparison of these two lines highlights the significant impact of incorporating H$_{2}$ cooling, as it brings the model predictions into much closer agreement with occupation fraction estimates inferred from observations (shown by the blue and yellow bands), particularly for dwarf galaxy formation in halos with M$_{\mathrm{halo}} < 2$--$3 \times 10^8$ M$_{\odot}$, corresponding to virial temperatures of approximately 10$^4$ K, below which the efficiency of atomic hydrogen cooling rapidly diminishes. 

Furthermore, we investigate the effects of incorporating two different background radiation models, \citetalias{2012ApJ...746..125H} and \citetalias{2020MNRAS.493.1614F}, as described in section \ref{sec:UVradiation}. The inclusion of UV background radiation suppresses the formation of H$_2$ in low-mass halos and so has an influence on the formation of dwarf galaxies, resulting in a shift of the occupation fraction predictions towards higher masses. The main difference between the two UV background models lies in the chosen redshift of reionization, after which UV background radiation suppresses H$_2$ formation in low-mass halos. In the case of \citetalias{2012ApJ...746..125H}, characterized by an earlier reionization redshift, we observe an earlier suppression of ultra-faint galaxy formation, thereby elevating the threshold for formation of galaxies in the occupation fraction results. We have confirmed that this result is almost entirely due to the difference in reionization redshifts between the FG20 and HM12 models, rather than, e.g., the spectral distribution of UV radiation. It is worth noting that effects of inhomogeneous reionization have not been explicitly considered in our model. Previous studies have shown that these inhomogeneities may lead to varying reionization times for low-mass halos in diverse environments \protect\citep{Katz2020, Ocvirk2021}, potentially introducing scatter in the predictions for occupation fractions.

%%%%%%%%%%%%%%%%
%%%%%%%%%%%%%%%%

In order to validate our results and provide a comprehensive comparison, we compare our findings with two independent studies. Firstly, we considered the forward-modeling framework for MW satellites presented by \cite{Nadler2020}. Their model extends the abundance-matching framework \citep{2018ARA&A..56..435W} into the dwarf galaxy regime by parametrizing the galaxy--halo connection---including the faint-end slope of the luminosity function, the galaxy--halo size relation, the scatter in galaxy luminosity and size, and the disruption of subhalos due to baryonic effects \citep{Nadler2018,2019ApJ...873...34N}---and constraining these parameters using recent MW satellite observations. In particular, \cite{Nadler2020} focused on MW satellites detected in photometric data from DES and PS1, which together cover a significant portion of the high Galactic latitude sky, including the contribution of satellites originally associated with the LMC. Importantly, they incorporated position-dependent observational selection effects that accurately represented satellite searches in imaging data from surveys such as DES and PS1. In our comparisons, we utilized their posterior on the galaxy occupation fraction, where the dark and light colors in Fig.~\ref{fig:occupationFraction} correspond to the 1 and 2 $\sigma$ confidence intervals, respectively, and the median is represented by the blue curve. We find that our most realistic model which incorporates H$_{2}$ cooling and utilizes the UV background radiation prescription from \citetalias{2020MNRAS.493.1614F}, lies within the $2\sigma$ uncertainty of the  occupation fraction inferred from observations by \cite{Nadler2020}.

Additionally, we examined the results obtained from the regulator-type modeling technique introduced in \cite{2022MNRAS.514.2667K} and employed by \cite{Manwadkar2022} to model the MW satellite population. This approach allowed for an exploration of the luminosity function by forward modeling observations of the population of dwarf galaxies while accounting for observational biases in surveys through their respective selection functions. Furthermore, they incorporated current constraints on the MW halo mass and the presence of the LMC. In our analysis, we utilized the shaded orange region on the plot, where the dark and light colors represent the $1$ and $2$ $\sigma$ dispersion, respectively, and the median is indicated by the orange curve.

By comparing our results with these complementary approaches, we find agreement within the 2 $\sigma$ dispersion range of the respective results. However, based on the median of our findings, we estimated that the peak mass above which 50\% of the halos host a luminous component is approximately a factor of 2 higher than the predictions by \cite{Nadler2020} and \cite{Manwadkar2022}. 

It is important to highlight that {\sc Galacticus} does not currently account for any pre-infall mass loss from halos. Nevertheless, N-body simulations demonstrate that peak masses are typically attained before infall, as the effects of tidal stripping begin to diminish the mass to some extent prior to infall \citep{Behroozi2014}. To account for these uncertainties, we include a shaded region representing a 40\% uncertainty in the determination of peak masses derived from our SAM prediction. The implementation of this missing physics is currently underway (Du \& Benson, in prep.) in {\sc Galacticus}.

As a result of this caveat, our current model likely overestimates peak masses due to the absence of accounting for pre-infall mass loss. With improved modeling in this regard, we anticipate our estimates to align more closely with these alternative models. Specifically, our estimate suggest that approximately 50\% of the halos with peak masses around $\sim 8.9 \times 10^{7}$ M$_{\odot}$ would host a luminous component, while \cite{Nadler2020} inferred a best-fit value of $\sim 4.2 \times 10^{7}$ M$_{\odot}$ and \cite{Manwadkar2022} predicted a value of $\sim 3.5 \times 10^{7}$~M$_{\odot}$. 

Comparing these findings against occupation fraction predictions from hydrodynamical simulations targeting similar halo mass ranges reveals that these simulations consistently predict a cutoff in ``galaxy formation'' at higher halo masses. In particular, many hydrodynamical predictions span a range of $6.5 \times 10^8$ to $3.5 \times 10^{9} \mathrm{M}_\odot$ for the bound mass at which $50\%$ of halos host galaxies, depending on the specific model configurations and reionization redshift assumptions employed \citep{Sawala2016focc, Benitez2017, Benitez2020}\footnote{Note that the work by \protect\cite{Benitez2020} analyze results based on both hydrodynamical and semi-analytical simulations.}. Importantly, it should be emphasized that the definition of occupation fraction in these simulations is subject to resolution limitations, and the effects of H$_2$ cooling must also be considered. These factors notably contribute to the disparities witnessed in the results. Nevertheless, due to the inherent dissimilarities in modeling approaches, a direct comparison between our SAM model and hydrodynamical simulations is not straightforward. For example, the simulation of \protect\cite{Agertz2019} forms a dwarf with $M_\star \approx 3 \times 10^4\mathrm{M}_\odot$ in a halo of mass approximately $8 \times 10^8\mathrm{M}_\odot$, while the simulation of \protect\cite{Applebaum2021} produces several galaxies in halos in the (bound) mass range $10^7$--$10^9\mathrm{M}_\odot$. These simulations are therefore consistent with our model (i.e. they imply that the occupation fraction is greater than zero at these halo masses), but do not allow for a detailed characterization of the cut-off in the occupation fraction, precluding a careful comparison with our results. For a more comprehensive understanding of how diverse assumptions and models can influence the predicted occupation fraction, please refer to Appendix~\ref{app:occ}.

\subsubsection{Stellar mass -- Halo mass relation}

\begin{figure}
	\includegraphics[width=\columnwidth]{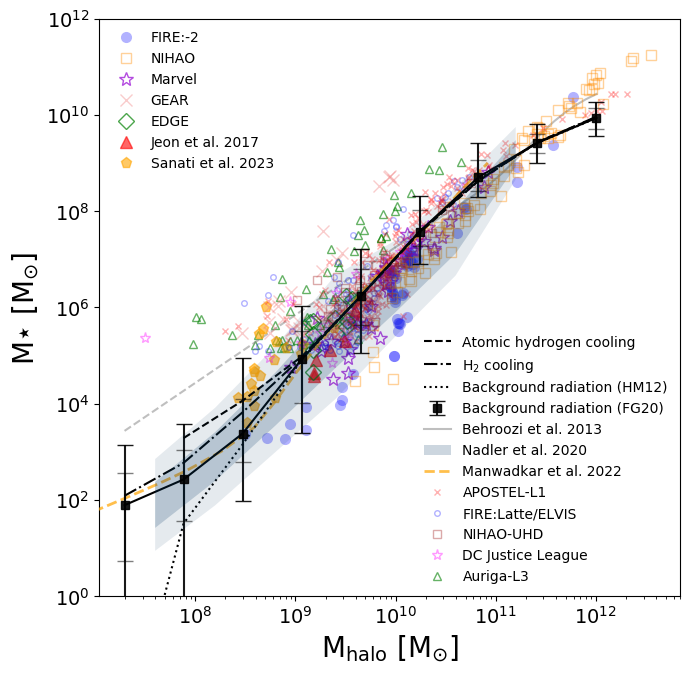}
    \caption{Stellar mass--halo mass relation. Our model predictions are represented by black curves with different line styles. We compare our results to the constrained/extrapolated SMHM relation from \protect\citealt{Behroozi_2013} (depicted by the grey curve/dashed grey curve), the results from \protect\citealt{Nadler2020} (illustrated by the shaded blue region), the results from \protect\citealt{Manwadkar2022} (shown by the orange dashed line), and simulations of central/field dwarf galaxies in MW-like environments from various models, as well as simulations that zoom-in on individual dwarf-mass halos (represented by different markers. Please refer to the text or see Figure 2 in \protect\citealt{Sales2022} for more details).}
    \label{fig:SMHM}
\end{figure}

Fig.~\ref{fig:SMHM} showcases the stellar mass--halo mass (SMHM) relation, which provides crucial insights into the connection between the masses of galaxies and their dark matter halos. We present the median values of the SMHM relation obtained from our model, incorporating the various physical processes discussed in section \ref{methods}. Each line style corresponds to a specific combination of physics, as outlined earlier (see section \ref{sec: occupation fraction}). Our most realistic model, which includes H$_{2}$ cooling and the UV background radiation prescription from \citetalias{2020MNRAS.493.1614F}, is represented by the error bars indicating the $1$ and $2$ $\sigma$ dispersion around the median value.

In terms of consistency with previous studies, our estimations for the higher mass end align well with a range of simulations and the abundance matching model by \cite{Behroozi_2013} as illustrated by the grey solid line. We show an extrapolation of that relation to lower mass systems in dashed gray, from which our results start to substantially deviate downwards for $M_{\rm halo}<10^9\; \rm M_\odot$. Notably, we find overall agreement with recent results from \cite{Nadler2020}, whose SMHM relation inferred from MW satellite observations is depicted by the shaded blue region, with darker and lighter shades corresponding to the $1$ and $2$ $\sigma$ confidence intervals, respectively. 

Additionally, we compare our results with available simulations of central/field dwarf galaxies in MW-like or Local Group-like environments (with data compiled by \citealt{Sales2022})\footnote{It is important to acknowledge that discrepancies might arise when comparing isolated dwarfs from hydrodynamical simulations due to potential variations in the definition of halo mass. Our model specifically focuses on dwarf satellites within MW systems. However, the purpose here is to emphasize the general concurrence between the outcomes of our model and the findings of existing simulations.}. For these comparisons, different markers are used, as indicated in the lower right part of the plot. The marker guide includes red crosses representing APOSTLE, L1 resolution \citep{Sawala2016, Fattahi2016}, blue open circles showing Latte and ELVIS suites \citep{Wetzel2016, Garrison-Kimmel2019} of FIRE-2 simulations\footnote{The latest version of FIRE simulation (FIRE-3) shows even better agreement with our predictions (see Figure 9 in \citealt{Hopkins2023}).} \citep{Hopkins2018}, brown squares representing NIHAO-UHD \citep{Buck2019}, pink stars showing DC Justice League \citep{Munshi2021}, green triangles representing Auriga, L3 resolution \citep{Grand2017}, while the legend in the upper left corner denotes simulations that zoom-in on individual dwarf-mass halos. These include blue circles showing FIRE-2 \citep{Fitts2017, Wheeler2019, Hopkins2018, Wheeler2015}, orange squares showing NIHAO \citep{Wang2015}, purple stars showing Marvel \citep{Munshi2021}, orange crosses showing GEAR \citep{Revaz2018}, green diamonds showing EDGE \citep{Rey2019, Rey2020}, red triangles showing work by \cite{Jeon2017}, and orange pentagons show results by \protect\cite{Sanati2023}.

The agreement observed with various simulations provides strong support for the validity of our modeling approach. Importantly, thanks to the use of SAMs, our predictions extend to fainter regimes, surpassing the capabilities of state-of-the-art hydrodynamical simulations. Overall, our different models comparing the effect of including various physics remain consistent with each other within the 2 $\sigma$ dispersion in the SMHM relation. However, some deviations are observed in the ultra-faint regime, where the model incorporating H$_2$ cooling and UV background radiation from \citetalias{2020MNRAS.493.1614F} produces the best results in terms of agreement with previous works. It is worth noting that our model slightly underpredicts the stellar mass content in the central galaxy (MW-analog). Nonetheless, the median value captures the lower end of stellar mass predictions for this halo mass range.

The SMHM relation predicted by {\sc Galacticus} unveils some intriguing features that align with findings from hydrodynamical simulations, such as the mass-dependent scatter in the SMHM relation, which exhibits an increasing trend around the median in the ultra-faint regime. This behavior seems to be influenced by the impact of formation histories, particularly the duration of star formation prior to reionization, directly affecting the stellar mass content at low redshifts \citep{Rey2019, Munshi2021}. Another interesting prediction emerges in the ultra-faint regime for $M_{\mathrm{halo}} < 10^9$ M$_{\odot}$ (corresponding to $M_{\star} < 10^5$ M$_{\odot}$), where the power-law relation in the SMHM appears to undergo a break. Remarkably, this feature appears to correlate with the dominance of H$_2$ cooling and is further amplified by the effects of UV background radiation, specifically the time of reionization. These predictions are consistent with the SMHM relation obtained from forward modeling results by \citeauthor{Manwadkar2022}~(\citeyear{Manwadkar2022}; depicted by the dashed orange line in Fig.\ref{fig:SMHM}), although the position of the break in this study reflects the inefficiency of supernova-driven winds in the smallest galaxies.

%%%%%%%%%%%%%%%%
%%%%%%%%%%%%%%%%

\subsubsection{Luminosity function}\label{sec:lum_function}

\begin{figure}
    \includegraphics[width=\columnwidth]{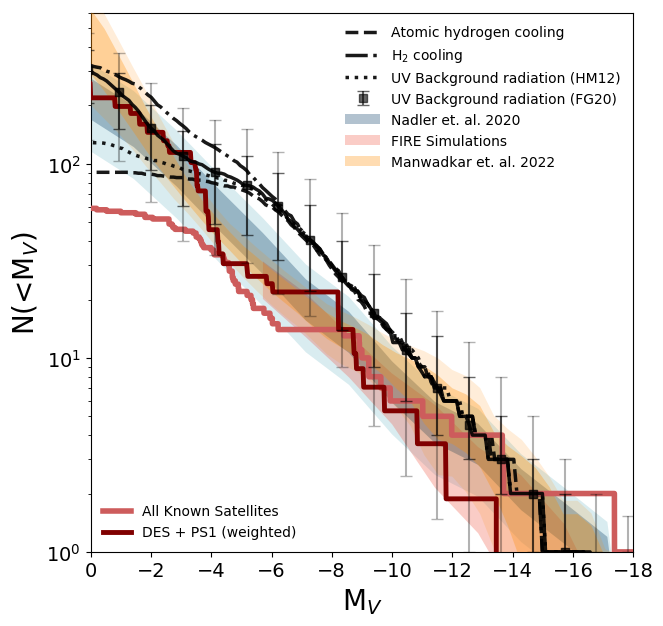}
    
    \caption{The luminosity function of MW satellites satisfying the criteria of M$_{V}$ $<$ 0, r$_{1/2}$ $>$ 10 pc, and a maximum distance of 300 kpc from the MW. Our model's predictions, represented by black curves with distinct line styles, are compared to observational data for all known MW satellites (light red curve) and the estimate derived in \protect\citealt{Drlica-Wagner2020} (maroon curve), which corrects for observational data incompleteness. Additionally, we present the results from simulations by \protect\citealt{Nadler2020} (blue shaded region), \protect\citealt{Manwadkar2022} (orange shaded region), and hydrodynamic simulations of the Local Group using the FIRE feedback prescription (pink shaded region) by \protect\citealt{Garrison-Kimmel2019}.}
    \label{fig:luminosityFunction}
\end{figure}

Fig.~\ref{fig:luminosityFunction} demonstrates our model's predictions for the luminosity function of the MW satellite population. To ensure consistency with conducted observations, we impose two selection criteria: satellites must reside within a distance of 300 kpc from the host halo's center, and they should have a minimum half-light radius of $r_\mathrm{h}$ $>$ 10 pc. The error bars on the plot represent the 1 and 2 $\sigma$ dispersion due to host-to-host scatter (across a range of halo masses). Our most accurate model predicts a total of  $300 ^{+75} _{-99}$ ($300 ^{+166} _{-170}$) satellites with an absolute $V$-band magnitude (M$_{V}$) less than 0, for 1$\sigma$ (2$\sigma$) dispersion.

By examining our models incorporating various physics components (similar line styles as Fig.~\ref{fig:occupationFraction} and Fig.~\ref{fig:SMHM}), we discern their impact on the resulting luminosity function. Notably, the inclusion of H$_{2}$ cooling leads to a considerable increase in the number of predicted ultra-faint satellites, surpassing a factor of $>$ 3, while the incorporation of UV background radiation serves to flatten the luminosity function at the ultra-faint end.

To assess the agreement with observational data, we compare our predictions with the luminosity function of all known MW satellites (light red curve) and the DES+PS1 data (from \citealt{Drlica-Wagner2020}), corrected for observational incompleteness (maroon line). It is important to note that the light red curve exhibits a more pronounced flattening at the ultra-faint end due to the incompleteness in the observations. In contrast, our results closely capture the rise predicted in the weighted DES+PS1 data (refer to \cite{Drlica-Wagner2020} for details of estimation), with the total number of satellites falling within the 2 $\sigma$ dispersion.

Examining the higher end of the luminosity function, we find agreement (within the 2 $\sigma$ dispersion) between our model and observational results (although we do not constrain our model to produce analogs of the LMC and SMC in all cases). However, it is worth emphasizing that the weighted DES+PS1 results do not encompass the LMC, SMC, and Sagittarius, accounting for the lower values observed compared to the all-known case at the higher end.

Moreover, we juxtapose our results with previous forward modeling methods, including the work by \cite{Nadler2020} (depicted by the blue shaded region) and \cite{Manwadkar2022} (illustrated by the orange shaded region) (as introduced in section \ref{sec: occupation fraction}). Additionally, we incorporate the FIRE hydrodynamical simulation by \cite{Garrison-Kimmel2019}, extending down to the FIRE resolution limit of $\sim -6$ mag (represented by the pink shaded region). These systems do not explicitly include analogs of the LMC or SMC. Overall, our results demonstrate strong agreement with previous simulations and forward modelling approaches, albeit with a slight tendency to overpredict the median number of satellites. Notably, in the ultra-faint regime, discrepancies arise between observational data and various simulations; however, the simulations generally converge within the 2 $\sigma$ limit. Remarkably, our best-performing model closely reproduces the predicted weighted DES+PS1 data at the low-mass end of the luminosity function.

In light of the higher median predicted for the satellite luminosity function in our model compared to other studies, such as \cite{Nadler2020}, it is important to consider some underlying differences of the respective models. For instance, variations in the underlying subhalo mass functions predicted by {\sc Galacticus} and cosmological zoom-in simulations (e.g., see Fig.~10 of \citealt{2023ApJ...945..159N}) may account for some of the discrepancy. Additionally, the extent to which dark matter subhalos are disrupted by the central galaxy could also influence the resulting luminosity functions. In our model, subhalos are tidally stripped using the \cite{Pullen2014} prescription, including the potential of the central galaxy, while  \cite{Nadler2020} apply a random-forest model trained on hydrodynamic simulations to capture this effect \citep{Nadler2018}.\footnote{We note that \cite{Hartwig2022} predict a total number of MW satellites comparable to our results without incorporating H$_2$ cooling or accounting for tidal stripping due to the central galaxy.} Importantly, our main results are robust in the sense that our predictions for the occupation fraction and SMHM relation do not change if we restrict to the subset of merger trees that produce luminosity functions similar to \cite{Nadler2020}. We leave direct calibration of our model based on forward-modeling the observed MW satellite population to future work.

%%%%%%%%%%%%%%%%
%%%%%%%%%%%%%%%%

\subsection{Dwarf population}\label{Dwarf population}

In this study, we utilize the optimal model presented in section \ref{methods}, which incorporates the physics of molecular hydrogen cooling, UV background radiation, and IGM metallicity. Our aim is to predict properties of the dwarf galaxy population and compare these to existing observations\footnote{Observational data are compiled from various resources (mainly from \citealt{Drlica-Wagner2020, Simon2019, McConnachie2012}). When multiple data sources exist for a given galaxy, we use the most precise and/or accurate measurement.} and simulations.

\subsubsection{Mass--metallicity relation}\label{sec:mass-metal}

\begin{figure}
\includegraphics[width=\columnwidth]{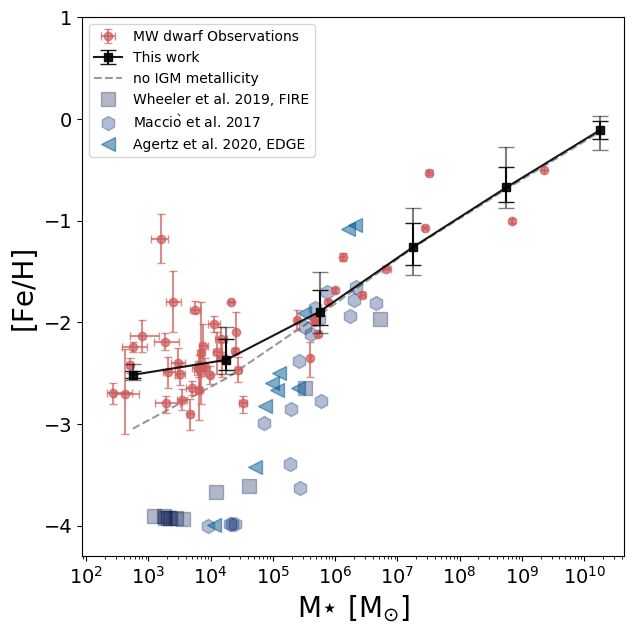  }
    
    \caption{Stellar mass--metallicity relation for galaxies. The black curve, along with the black and grey error bars, represents the median value, and the 1 and 2 $\sigma$ dispersions, respectively, derived from our simulation's predictions. The gray dashed line represent the predictions of our model without the IGM metallicity included. Blue markers indicate results from hydrodynamical simulations (\protect\citealt{Wheeler2019}: blue squares; \protect\citealt{Maccio2017}: blue hexagon; and \protect\citealt{Agertz2019}: by blue triangles). Red markers with error bars depict the observational results for dwarf galaxies located within 300 kpc of the MW, compiled primarily from studies by \protect\citealt{Drlica-Wagner2020, Simon2019, McConnachie2012}.}
    \label{fig:mass-metal}
\end{figure}

The metallicity of a galaxy is commonly quantified by the iron to hydrogen abundance ratio ([Fe/H]). As shown in Fig.~\ref{fig:mass-metal}, we present the mean stellar [Fe/H]\footnote{In our present Galacticus model, [Fe/H] is computed using the instantaneous recylcing approximation, and the assumption of Solar abundance ratios.} as a function of stellar mass ($M_\star$) for our simulation. The black curve represents the median value, while the black and grey error bars denote the $1$ and $2 \; \sigma$ dispersion, respectively. To validate our results, we compare them with observations of dwarf galaxies located within a 300 kpc radius of the MW (illustrated by red markers). The observations indicate the presence of a metallicity plateau around [Fe/H]$\sim -2.5$, which is reproduced well by our simulation incorporating the IGM metallicity model.

%%%%%%%%%%%%%%%%
%%%%%%%%%%%%%%%%

Interestingly, the mass-metallicity relation for the very low-mass satellites appears to be strongly influenced by the evolution of IGM metallicity as a function of redshift. This influence becomes apparent when comparing the black curve, which includes IGM metallicity in our model, with the dashed grey curve, where the IGM metallicity is excluded, and which shows a power-law extension to low masses with no plateau\footnote{With no IGM metallicity, the metallicities of our galaxies are determined by our feedback/outflow model, which has a simple power-law dependence on halo mass, and so necessarily leads to a power-law mass-metallicity relation.}. (The inclusion of IGM metallicity significantly affects the predicted metallicities of these satellites---essentially setting a floor in metallicity corresponding to the metallicity of the IGM gas accreted at the time at which the galaxy formed---highlighting the importance of the surrounding cosmic environment in shaping their chemical enrichment history.) When comparing our findings to zoom-in hydrodynamical simulations (such as those conducted by \citealt{Agertz2019, Wheeler2019, Maccio2017}), we find that these simulations tend to predict near-primordial abundances for objects with stellar masses below $10^5$ M$_{\odot}$. However, it is important to note that the examples presented in this study do not have a large cosmological environment and thus are not enriched by nearby sources (for a comprehensive comparison with recent simulation predictions refer to Figure 1 in \citealt{Sanati2023}). The implications of this lack of enrichment (in hydrodynamic simulations) remain uncertain and necessitate further investigation. 

Recent studies have considered a few possible self-consistent avenues to populate the plateau in [Fe/H] at the faintest end of the mass–metallicity relation.
The study by \protect\cite{Prgomet2022}, using the adaptive mesh refinement method, studied the effect of varying the IMF on the evolution of an ultra-faint dwarf. In this framework, at low gas metallicities, the IMF of newborn stellar populations becomes top-heavy, increasing the efficiency of supernova and photoionization feedback in regulating star formation. The increase in the feedback budget is none the less met by increased metal production from more numerous massive stars, leading to nearly constant iron content at $z = 0$ that is consistent with the results achieved from our model (for their case at a stellar mass of $M_{\star} = 10^3\mathrm{M}_\odot$, the typical metallicity is [Fe/H] $\sim -2.5$). Additionally, the study by \protect\cite{Sanati2023}, running zoom-in chemo-dynamical simulations of multiple halos and including models that account for the first generations of metal-free stars (Pop III), demonstrate an increase in the global metallicity of ultra-faints, although these are insufficient to resolve the tension with observations (see their fig.~6).

Several studies have examined the effect of different feedback processes on shaping the dwarf population (see, for example, \protect\citealt{Lu_2017, Agertz2019, Smith2021}).
In this context, the work by \cite{Lu_2017} using a semi-analytical model provides valuable insights. Their findings shed light on the connection between preventive and ejective feedback mechanisms and the stellar mass function and mass-metallicity relation of Milky Way dwarf galaxies. Where preventive feedback acts to inhibit baryons from accreting onto galaxies, and in the realm of low-mass halos, a commonly employed form of preventive feedback in SAMs is photoionization heating. This mechanism effectively reduces radiative cooling and mass accretion in low-mass halos, thereby influencing the evolution of these galaxies. On the other hand, ejective feedback processes involve the expulsion of baryons from the galaxy into the IGM, often characterized by the presence of outflows. These mechanisms play a significant role in shaping the gas content and subsequent star formation in dwarf galaxies. By incorporating both preventive and ejective feedback in their model, \cite{Lu_2017} demonstrate the ability to simultaneously match the observed stellar mass function and the mass-metallicity relation. Moreover, they highlight the importance of considering a redshift dependence for preventive feedback, although the precise nature of this dependence remains largely uncertain.

Building upon the insights from \cite{Lu_2017}, our results further support the notion that the mass-metallicity relation for low-mass dwarfs is intricately linked to the interplay between feedback processes and the enrichment of the surrounding environment (i.e. enrichment of the IGM). We acknowledge that our approach is not self-consistent, as we do not explicitly account for the metal outflows from our galaxies and their mixing into the IGM. However, the inclusion of IGM metallicity in our model becomes imperative to achieve consistency with observational data, as demonstrated by our agreement with observations.

Another study, conducted by \cite{Pandya2021}, showcases that the mass loading factors for winds in dwarf galaxies can be large (i.e. $\gg1$; as evident from their Fig. 7), and these winds are responsible for carrying away a significant portion of the produced metals. They also reveal that higher mass galaxies exhibit substantially lower mass loading factors for their winds, along with lower metal-loading factors. This finding suggests that dwarf galaxies may play a substantial role in enriching the IGM. Given these compelling facts, our SAM approach has the potential to allow us to resolve the dwarf galaxies and accurately predict IGM metal enrichment. Simultaneously, our SAM enables us to model the massive halos, which actively accrete gas from the enriched IGM, facilitating a comprehensive understanding of the intricate interplay between galaxies and their surrounding environment.

\subsubsection{Size--mass relation} \label{size-mass}

\begin{figure}
    \includegraphics[width=\columnwidth]{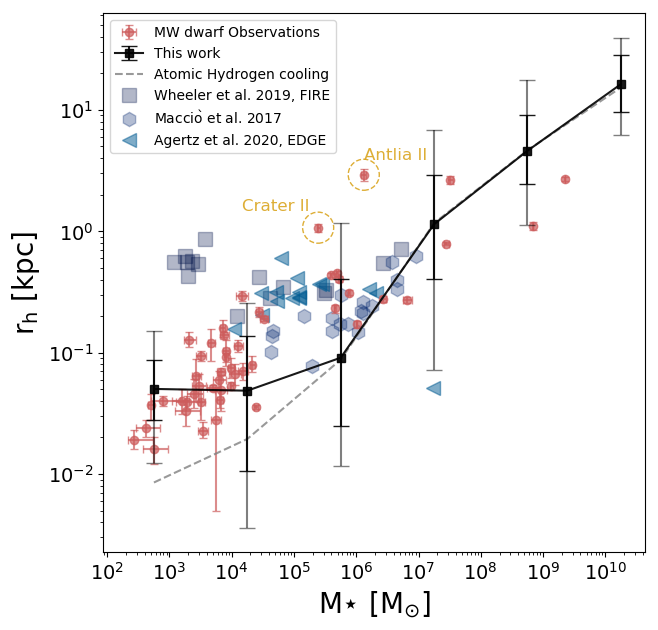}
    \caption{The size (projected half stellar mass radius)--stellar mass relation for dwarf galaxies. The black curve, along with the black and grey error bars, represents the median value, and the 1 and 2 $\sigma$ dispersions, respectively, derived from our simulation's predictions. The gray dashed line represent the predictions of our model only including atomic hydrogen cooling. Blue markers demonstrate results from hydrodynamical simulations (\protect\citealt{Wheeler2019}: blue squares; \protect\citealt{Maccio2017}: blue hexagon; and \protect\citealt{Agertz2019}: blue triangles). Red markers with error bars depict the observational results for dwarf galaxies located within 300 kpc of the MW, compiled primarily from studies by \protect\citealt{Drlica-Wagner2020, Simon2019, McConnachie2012}. }
    \label{fig:size-mass}
\end{figure}

We measure the projected half mass radius ($r_\mathrm{h}$) for all galaxies in our sample and plot it against the predicted stellar masses. As depicted in Fig.~\ref{fig:size-mass}, the black curve represents the median value, while the black and grey error bars indicate the $1$ and $2$ $\sigma$ dispersion, respectively. Our predictions successfully capture the size-mass relation for the majority of observed galaxies (depicted by red markers) within the 2 $\sigma$ range of our sample. Interestingly, we find that systems resembling Antlia II and Crater II are sometimes predicted by our model, although they lie far away from the median of the relation predicted by the model. Such galaxies correspond to the high angular momentum tail of the distribution of galaxy angular momenta---we will discuss the relation between size and angular momentum in more detail below. When comparing our results to hydrodynamical simulations, we generally agree with their best predictions above the $\sim$ $10^{5}$ M$_{\odot}$ limit, with the exception of a few extreme cases (e.g., the outlier presented by \cite{Agertz2019} where no feedback is included).

%%%%%%%%%%%%%%%%
%%%%%%%%%%%%%%%%

In our simulation, sizes are determined by the specific angular momentum content of stars and gas, as described by the equation:

\begin{equation}
    j = v_\mathrm{h}  r_\mathrm{h} = (\mathrm{G} M_\mathrm{h}/r_\mathrm{h})^{1/2} r_\mathrm{h} = (\mathrm{G} M_\mathrm{h} r_\mathrm{h})^{1/2},
\end{equation}
where $v_\mathrm{h}$ is the rotational speed at the half mass radius, $r_\mathrm{h}$ is the half mass radius, and $M_\mathrm{h}$ is the total mass content within the half mass radius. Given that intermediate and low-mass dwarfs are predominantly dark matter-dominated, and we have a reasonably accurate SMHM relation and a correctly modeled occupation fraction distribution, it is likely that the dark matter mass estimate is accurate. If we aim to explain the changes of slope in the size--mass relation of galaxies, the most apparent approach would be to look at the changes in the angular momentum content.

The angular momentum is primarily determined by the angular momentum of the gas in the halo during its formation, and subsequently, by the fraction of that angular momentum that is transferred into the galaxy through cooling and gas accretion, as well as the fraction that is expelled by outflows. These factors encompass a certain level of uncertainty. In our current model, we address the inefficiencies of atomic hydrogen cooling by incorporating H$_2$ cooling. Specifically, for temperatures below $10^4$ K, corresponding to halo masses around $10^9 \mathrm{M}_\odot$, which host galaxies with stellar mass components ranging from $10^4$--$10^5 \mathrm{M}_\odot$, the dominant cooling mechanism becomes H$_2$ cooling. Additionally, we include the UV background radiation model by \citetalias{2020MNRAS.493.1614F}, which suppresses gas accretion. From Fig.~\ref{fig:SMHM}, we observe that its effects are maximized for dwarfs with stellar masses below $10^5 \mathrm{M}_\odot$. The overall effect becomes evident when comparing the black solid line representing our optimal model to the dashed grey line, where only atomic hydrogen cooling is present and no UV background radiation was used. These results suggest that variations in cooling mechanisms along with gas accretion suppression can account for the observed changes in the slope at these particular mass scales. TAR

\subsubsection{Velocity dispersion---mass relation}

\begin{figure}
    \includegraphics[width=\columnwidth]{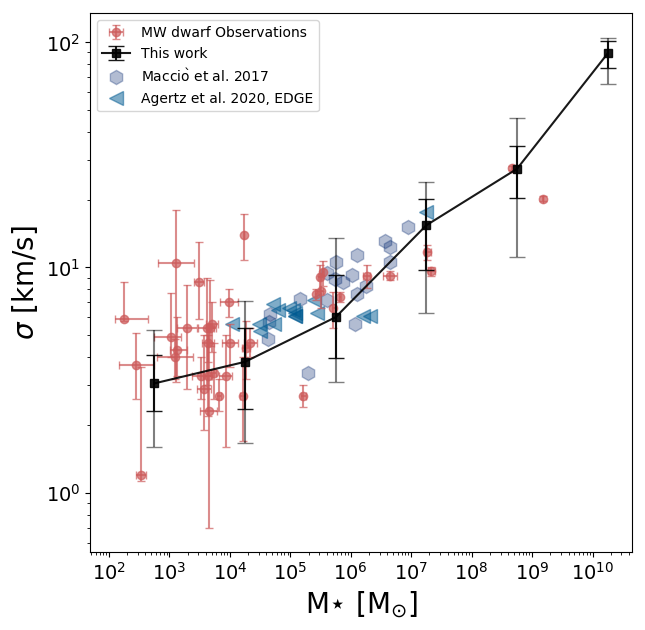}
    \caption{The 1D line-of-sight velocity dispersion (measured at the half stellar mass radius)--stellar mass relation. The black curve, along with the black and grey error bars, represents the median value, and the 1 and 2 $\sigma$ dispersions, respectively, derived from our simulation's predictions. The hydrodynamical simulation results are shown by blue markers, with \protect\citealt{Maccio2017} represented by blue hexagons, and \protect\citealt{Agertz2019} by blue triangles. The red markers with error bars depict the observational results for dwarf galaxies located within 300 kpc of the MW, compiled primarily from studies by \protect\citealt{Drlica-Wagner2020, Simon2019, McConnachie2012}. Our results demonstrate agreement with the velocity dispersion--mass relation in higher mass galaxies, while indicating lower median predictions for galaxies with stellar masses below $10^{5}$ M$_{\odot}$.}
    \label{fig:velDispersion-mass}
\end{figure}

We measure the 1D line-of-sight velocity dispersions at the half stellar mass radius for all galaxies in our sample and plot them against the predicted stellar masses. In Fig.~\ref{fig:velDispersion-mass}, similar to Fig.~\ref{fig:size-mass}, the black curve represents the median value, while the black and grey error bars indicate the $1$ and $2$ $\sigma$ dispersion, respectively. Our predictions successfully reproduce the velocity dispersion--mass relation for observed galaxies within the 2~$\sigma$ limit of our sample (all the observational data are represented by red markers). We compared our results with hydrodynamical simulations by \cite{Maccio2017, Agertz2019}, shown by blue markers, finding general agreement within the 2~$\sigma$ dispersion limit. 

It is worth noting that our model does not fully capture the observed scatter in 1D velocity dispersions at the lower mass end. Several potential reasons may explain this. Firstly, it is possible that our current model does not incorporate all the relevant physical processes that govern the ultra-faint regime. The intricate dynamics and feedback mechanisms specific to these low-mass galaxies could play a significant role in shaping their velocity dispersions. Secondly, observational limitations introduce additional uncertainties in our measurements. Factors such as contamination from foreground stars in the MW and the influence of binary stars within the sample of stars from the ultra-faint dwarfs (see \citealt{Simon2019} for further details) could contribute to the observed large dispersions. 

We would like to highlight that, given the observational uncertainties, our model's predictions align well with the data, providing consistency without necessitating the inclusion of core formation. However, it is crucial to emphasize that these observational uncertainties also mean that we cannot conclusively rule out the possibility of core formation being present. This highlights the need for improved and more precise measurements in order to better understand and constrain the underlying physical processes. Additionally, our model's success in matching the velocity dispersion, combined with accurate predictions of the occupation fraction, suggests that it is effectively free of the too-big-to-fail problem \citep{Boylan2011}.

%%%%%%%%%%%%%%%%
%%%%%%%%%%%%%%%%

\subsection{Mass function predictions for various halo masses}\label{Mass function predictions}

%%%%%%%%%%%%%%%%
\begin{figure*}
    \includegraphics[width=\columnwidth]{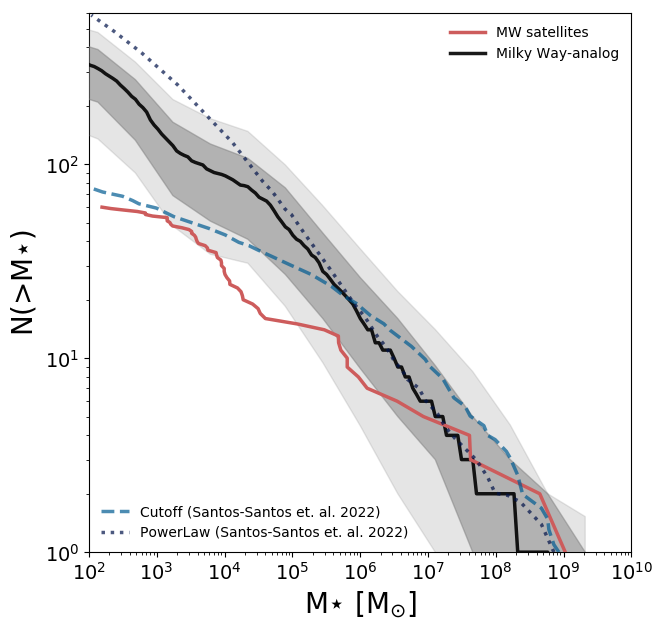}
    \includegraphics[width=\columnwidth]{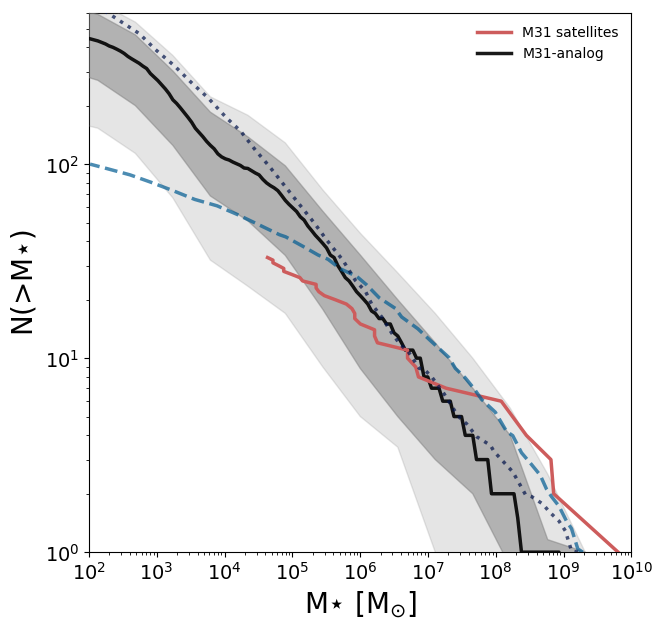}
    \includegraphics[width=\columnwidth]{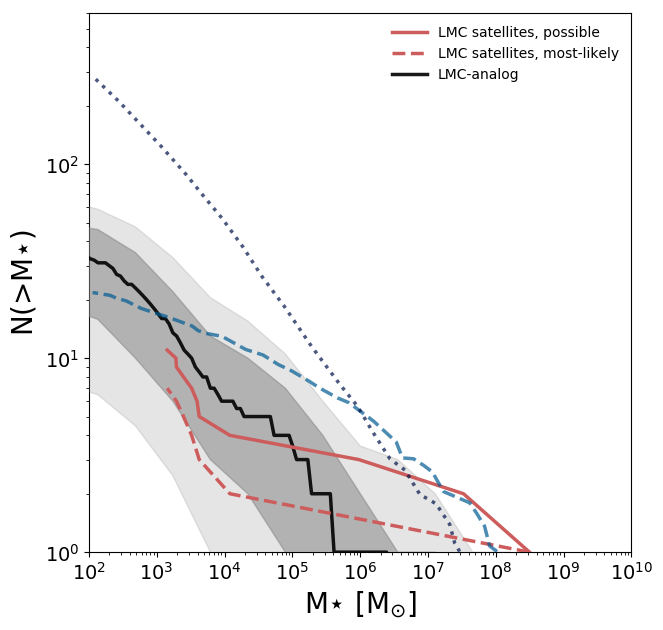}
    \includegraphics[width= 8.4cm, height= 8.15cm]{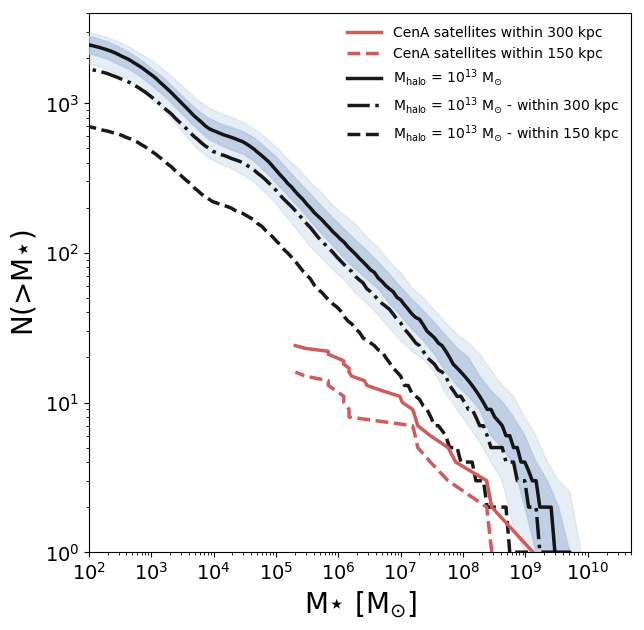}
    \caption{Predictions of our model for the cumulative stellar mass function of satellites. The black curve represents the median of our results, while the light and dark shaded regions indicate the 1 and 2 $\sigma$ dispersions, respectively. Observational constraints, if available, are shown by the red curves. The dashed and dotted blue lines correspond to the "Cutoff" and "PowerLaw" models from \protect\citealt{Santos-Santos2022}, allowing for a comparison with their results. Each panel displays our results for a different halo mass: the top left panel corresponds to the MW-analog halo, the top right panel to the M31-analog halo, the bottom left panel to the LMC-analog halo, the bottom right panel to a group-size halo with a mass of $10^{13}$ M$_{\odot}$.}
    \label{fig:massFunctions}
\end{figure*}
%%%%%%%%%%%%%%%%

Once calibrated, we can use our model to make predictions on the abundance of satellite galaxies for host systems with varying virial masses. In Fig.~\ref{fig:massFunctions}, we depict the cumulative stellar mass functions for subhalos associated with various halos of different masses, specifically showcasing satellites with stellar masses ($M_\star$) greater than $10^2 \rm \mathrm{M}_\odot$ and half mass radius ($r_{1/2}$) larger than 10 pc. The dark and light shaded grey regions represent the $1$ and $2$ $\sigma$ dispersion, respectively, while the black line shows the median of the results. For comparison, the red curve represents available observational results\footnote{The V-band magnitudes for satellites of the MW, M31, and LMC are sourced from \cite{McConnachie2012}'s revised compilation of Local Group dwarfs. For the Cen A system, values are extracted from \cite{Crnojevic2019}. Subsequently, the stellar masses are computed by employing relevant mass-to-light ratios derived from {\textsc Galacticus} predictions specific to the respective stellar masses.}, and the blue dashed and dotted curves represent the results from the abundance matching study by \cite{Santos-Santos2022}. The blue dotted line corresponds to their ``power-law'' model, assuming a power law relation for the $M_\star$ -- $V_\mathrm{max}$ relation, while the blue dashed curve corresponds to their ``cutoff'' model, assuming a cut-off in this relation.

In the top left panel, we present our results for the MW-analog. We ran 100 halos with virial masses ranging from $7 \times 10^{11}$ to $1.9 \times 10^{12} \rm \mathrm{M}_\odot$, in agreement with the current available mass constraints \citep{Wang2020, Callingham2019}. Our results show a reasonable agreement with the observations for the stellar masses of larger satellites (within 300 kpc from the MW). However, for the lower mass range, the discrepancy between our results and the observations becomes more prominent. This discrepancy could be partially attributed to incompleteness in the observational results, as we discussed in section \ref{sec:lum_function}, where estimations for corrections in the observational data predict much higher values for the number of MW satellites \citep{Tollerud2008, Drlica-Wagner2020}. Overall, our model suggests that only $\sim$20\% of the MW satellites with M$_V$ $<$ 0 have been discovered.

Comparing with the results of \cite{Santos-Santos2022}, we find reasonable agreement up to a stellar mass of $10^5 \rm \mathrm{M}_\odot$ for the satellites, and our results deviate from their predictions in the ultra-faint regime. Notably, the slope of our results in this regime shows better agreement with their power-law model, although the total predicted number of halos is a factor of $\sim$2 lower. It is worth mentioning that this slope was only achieved by including the effects of H$_2$ cooling, as our model with only atomic hydrogen physics shows flatter slopes, in better agreement with their cut-off model (see Fig.~\ref{fig:luminosityFunction} for comparison of our models).

The top right panel presents our results for the M31-analog. Similar to the MW case, we ran 100 realizations of a halo with a mass of $1.8 \pm 0.5 \times 10^{12} \rm \mathrm{M}_\odot$, in agreement with M31 mass constraints (from \citealt{Benisty2022, Shull2014, Diaz2014, Karachentsev2014}). Our results show agreement with the observations within the 2 $\sigma$ limit (albeit we get lower results for the higher mass end), although the surveyed population in M31 does not extend as deeply as our predictions show. Similar to the MW case, we observe agreement with \cite{Santos-Santos2022}'s results in the higher mass regime, while in the ultra-faint regime, our model predicts results closer to their power-law model. It is worth mentioning that their models assume an occupation fraction of 1 for their halos, whereas we found in section \ref{sec: occupation fraction} that only a fraction of our halos with peak masses below $2 \times 10^8 \rm \mathrm{M}_\odot$ host a luminous component.

The bottom left panel shows our results for the LMC-analog halo. We present the mass function results for satellite stellar masses based on 100 realizations of hosts with halo masses of $1.88 \pm 0.35 \times 10^{11} \mathrm{M}_\odot$ \citep{Shipp2021}. Our findings estimate that an isolated LMC-analog is expected to have approximately 33$^{+14}_{-12}$ satellites (for 1 $\sigma$ dispersion) with stellar masses above $10^2 \mathrm{M}_\odot$ and r$_{1/2}$ larger than 10 pc, lying within the halo's virial radius. Most of the realizations indicate that satellites have stellar masses below $4 \times 10^6 \mathrm{M}_\odot$, and the likelihood of generating an SMC within the virial radius is relatively low.

We then compare these results with the satellites associated with the LMC based on the kinematic analysis conducted by \cite{Santos-Santos2021}. According to their analysis, 11 of the MW satellites appear to have some connection with the LMC (``possible''), and from those, 7 show firm association (``most likely''). Our results seem to under-predict the number of higher mass subhalos while over-predicting the number of currently observed ultra-faint satellites. Apart from considering the effects of observational incompleteness, other factors may be at play here. Firstly, we do not constrain our LMC-analogs to have any high mass satellites such as the SMC. The occurrence of reproducing such a massive companion for the LMC in our model is probabilistically low, as only 1 such satellite was produced in our 100 realizations of the LMC, and it is located at a distance of approximately 140 kpc from the LMC-analog (beyond the virial radius of this halo/radius of approximately 100 kpc where we measure the associated satellites). Additionally, we are running our LMC-analogs as isolated halos and not in association with a larger halo such as the MW. The presence of a larger gravitational potential can more effectively disrupt the ultra-faint satellites, thereby decreasing the number of predicted satellites associated with the LMC. 

Comparing with the results from \cite{Nadler2020}, they predict 48 $\pm$ 8 LMC-associated satellites with $M_V < 0$ mag and $r_{1/2} > 10$ pc, approximately consistent with our predictions of 33$^{+14}_{-12}$ for 1$\sigma$ (33$^{+28}_{-26}$ for 2$\sigma$). This is also in reasonable agreement with the $\sim$70 satellites with $-7 < M_V < -1$ predicted by \cite{Jethwa2016} via dynamical modeling of the Magellanic Cloud satellite population. Additionally, our predictions can be compared with the work of \cite{Dooley2017}, who explored the satellite population of LMC-like hosts using several abundance-matching models and estimated $\sim$8-15 dwarf satellites with $M_* \geq 10^3 \rm \mathrm{M}_\odot$ within a 50 kpc radius of their hosts. Applying similar selection criteria to our model results gives us an estimation of 6$^{+6}_{-4}$. Furthermore, our results align well with the study by \cite{Jahn2019}, where they used five zoom-in simulations of LMC-mass hosts (with halo masses ranging from $1 \times 10^{11}$ to $3 \times 10^{11} \mathrm{M}_\odot$) run with the FIRE galaxy formation code, predicting $\sim$ 5--10 ultra-faint companions for their LMC-mass systems that have stellar masses above $10^4 \mathrm{M}_\odot$ (compared to our estimation of 6$^{+5}_{-4}$ for 1$\sigma$ dispersion). However, it is worth noting that our stellar mass function is steeper than their results. 

The bottom right panel illustrates our model's prediction for the satellite stellar mass function of subhalos within group-sized halos. These results are based on 100 realizations of a host halo with a mass of $10^{13} \, \mathrm{M}_{\odot}$. The shaded region in this plot is depicted in a distinct color as it differs from the other panels. In this case, the dispersion only represents variations resulting from constructing merger trees for the exact same halo mass, while in other panels, we include a range of masses for the halos, leading to a larger halo-to-halo scatter. 

Regarding the predicted stellar mass function for the satellites, we find more massive satellites compared to those in the MW and M31, along with a larger number of total subhalos (with $r_{1/2} > 10$ pc) within a radius of 450 kpc from the central galaxy (or the estimated virial radius). This trend is consistent with expectations for a halo with a larger virial mass. As a candidate in the nearby universe, we compare our results to Centaurus A (Cen A for short) with virial mass estimations ranging from $6.4 \times 10^{12}$ to $1.8 \times 10^{13} \mathrm{M}_\odot$ \citep{Karachentsev2007, vanDenBergh2000}. The V band magnitudes of Cen A's satellites were compiled from \cite{Crnojevic2019}. 

The study by \cite{Crnojevic2019} covers approximately half of the virial radius estimated for Cen A and includes satellites down to $M_V = -7.8$ (equivalent to a stellar mass of approximately $10^5 \mathrm{M}\odot$). Additionally, \cite{Crnojevic2019} provides results from earlier studies of Cen A \citep{Sharina2008, Karachentsev2013}, which target a wider region around the central galaxy, albeit with a lower limiting magnitude. Since the observational surveys each cover part of this group, we have adjusted the radius within which we make the comparison accordingly. Our model predictions depicted by the black dashed line, corresponds to satellite mass function within a radius of 150 kpc from the central galaxy. This selection mirrors the observational results with the same cuts, as shown by the red dashed line. Additionally, our results shown by the dashed-dotted line represent the satellite mass function within 300 kpc from the central galaxy, which can be compared to the observational data with similar cuts, as indicated by the red line.

Our results align well with the slope of the observational satellite stellar mass function at the higher mass end, although the exact number of predicted satellites is slightly higher. This can be interpreted as our results favoring a virial mass for Cen A close to the lower end of the current estimates, as number of satellites tend to scale on host halo mass. In any case, if we assume that a $10^{13} \mathrm{M}_\odot$ halo is a good representation of this system, our results suggest that a factor of $\sim$5-7 satellites with stellar masses above $10^5 \mathrm{M}_\odot$ are waiting to be discovered in this system. Additionally, a forthcoming study by Weerasooriya et al. (in prep.), utilizing the model outlined in \cite{Weerasooriya2022}, has also examined the Cen A system. Their prediction for the total count of satellites with $M_V$ magnitudes lower than -7.4 (equivalent to stellar masses around $10^{5} \mathrm{M}_\odot$) amounts to median number of 50. While the median is slightly lower than their compiled observational data for satellites of Cen A within 150 kpc\footnote{Their work encompasses a comprehensive compilation of the luminosity function for Cen A, including available observational data from \cite{Lauberts1989, deVaucouleurs1991, Karachentsev2003, James2004, Doyle2005, Sharina2008, Karachentsev2013, Muller2015, Muller2017, Muller2019, Crnojevic2014, Crnojević2016, Crnojevic2019, Taylor2016}. It is important to note that their dataset includes all dwarf candidates, not exclusively confirmed cases.}, the predicted distribution of the number of satellites still falls within the observed range. These results are marginally lower than our predictions below $M_V$ $\sim$ -10.

\section{Conclusions}\label{Conclusions}

In this study we have modified the {\sc Galacticus} semi-analytic model to incorporate key physical processes relevant to the formation of dwarf galaxies, and utilized that model to explore predictions for the galaxy-halo connection and the properties of the dwarf galaxy population of the Milky Way. Through the inclusion of essential physical processes such as IGM metallicity, H$_2$ cooling, and UV background radiation, coupled with the fine-tuning of various parameters, we have achieved significant success in replicating several characteristics observed in the dwarf galaxy population.

First and foremost, we find that our model with updated physics is able to reproduce the inferred SMHM relation while simultaneously reproducing the main physical properties of the dwarf galaxy population. This finding underscores the robustness of our model and its ability to capture the relationship between the stellar content and the underlying dark matter halos. Furthermore, our results demonstrate that the inclusion of H$_2$ cooling and a UV background radiation (prescribed by \citetalias{2020MNRAS.493.1614F}), motivated by recent observational constraints, is crucial to achieving an occupation fraction consistent with previous inferences. Our study reveals that the fraction of subhalos hosting galaxies with an absolute $V$-band magnitude less than 0 drops to 50\% at a halo peak mass of $\sim 8.9 \times 10^{7}$ M$_{\odot}$. Notably, earlier estimations based on older UV background estimates (\citetalias{2012ApJ...746..125H}) do not yield the same level of agreement. 

When examining the statistical properties of the MW dwarf population, we find broad success in reproducing key characteristics. Our predictions for the luminosity function of the MW dwarfs align well with observations once we account for the inherent halo-to-halo scatter. Remarkably, the presence of H$_2$ cooling is vital for capturing the large number of ultra-faint dwarf galaxies, highlighting its role in driving their formation. Our model predicts a total of $300 ^{+75} _{-99}$ satellites with an absolute $V$-band magnitude less than 0 within 300 kpc from our MW-analogs. This number would drop down to $91 ^{+42} _{-34}$ if we were to use our model including only the atomic hydrogen cooling. Our model of H$_2$ formation/destruction remains quite simplistic. Plausible changes in the underlying assumptions in computing metal cooling and H$_2$ formation/destruction under a radiation field (e.g. considering radiation from local sources, not just a mean background), could result in changes to the cooling efficiencies in small, early-forming halos. The efficiency and relevance of H$_2$ cooling in such halos remain subjects of ongoing debate (refer to Section 4.3.2 of the review by \citealt{Klessen2023}, and references therein).

Moreover, the inclusion of IGM metallicity enables us to successfully reproduce the mass-metallicity relation without the need for preventive feedback mechanisms. Our model achieves successful agreement with the sizes and velocity dispersions of ultra-faint dwarfs.

Finally, our model successfully predicts the stellar mass function of satellites for both MW and M31 analogs. Additionally, we use our model to make predictions for the two different mass scales: LMC and Cen A analogs. Our results demonstrate a general agreement with the available observational data, emphasizing the robustness of our model in generating predictions across a broad range of halo masses. The combined functionalities of this model, along with its comprehensive approach to predicting various aspects of the dwarf population, makes it uniquely powerful for investigating the faintest galaxy population across a range of environments/halo masses.

Looking ahead, there are several exciting directions to explore. Investigating how our results are influenced by the inclusion of an LMC-analog in the MW mass halos will provide valuable insights into the impact of satellite galaxies on the MW dwarf population, e.g. following the constrained merger tree methodology presented in \cite{2023MNRAS.521.3201N}. Furthermore, exploring alternative non-CDM models, such as self-interacting dark matter, will allow us to gauge the extent to which observations of dwarfs can inform our understanding of the nature of dark matter itself.

\section*{Acknowledgements}

We gratefully acknowledge the contributions of several individuals who supported this research. Special thanks to our referee, Martin Rey, whose valuable suggestions greatly improved this work. We would like to express our appreciation to Isabel Santos-Santos for generously providing the valuable data from \citep{Santos-Santos2022}. We are also thankful to Ana Bonaca, Viraj Manwadkar, Yves Revaz, Risa Wechsler, and Fakhri S. Zahedy for engaging in insightful conversations and offering valuable input throughout the course of this study. NA and LVS acknowledge financial support from NSF-CAREER-1945310 and NSF-AST-2107993 grants. NA would like to acknowledge the support provided by the UCR-Carnegie Graduate Fellowship. The computations presented here were conducted through Carnegie's partnership in the Resnick High Performance Computing Center, a facility supported by Resnick Sustainability Institute at the California Institute of Technology. Additional computations were carried out on resources made available through a generous grant from the Ahmanson Foundation.

%%%%%%%%%%%%%%%%%%%%%%%%%%%%%%%%%%%%%%%%%%%%%%%%%%
\section*{Data Availability}

The SAM model utilized in this project is openly accessible and can be accessed at the following link: \href{https://github.com/galacticusorg/galacticus/commit/b60e818869ea0bad7e2fcc2b9320cabbe02cf550}{https://github.com/galacticusorg/galacticus/}. Researchers and interested parties can freely explore and utilize the SAM model to replicate and build upon the findings presented in this study.

\bibliographystyle{mnras}
\bibliography{example} 

%%%%%%%%%%%%%%%%%%%%%%%%%%%%%%%%%%%%%%%%%%%%%%%%%%

%%%%%%%%%%%%%%%%% APPENDICES %%%%%%%%%%%%%%%%%%%%%

\appendix

\section{Details on constraining model}\label{app:model}

In this work, we utilize a model similar to that recently proposed by \cite{Weerasooriya2022}, but with some differences which result from the recalibration of our model after including the physics described in \S\ref{methods}. Here we outline the parameters that need adjustment.

\textit{Cooling rate: } We follow \cite{White-Frenk1991} to account for the cooling rates based on the following expression.
\begin{align}
    \dot{M}_\mathrm{cool} = \left\{ \begin{matrix} 4 \pi r_\mathrm{infall}^2 \rho(r_\mathrm{infall}) \dot{r}_\mathrm{infall} & \hbox{ if } r_\mathrm{infall} < r_\mathrm{hot, outer} \\ M_\mathrm{hot}/\tau_\mathrm{halo, dynamical} & \hbox{ if } r_\mathrm{infall} \ge r_\mathrm{hot, outer} \end{matrix} \right. ,
\end{align}

where $r_\mathrm{infall}$ is the infall radius in the hot halo and $\rho(r)$ is the density profile of the hot halo.

\textit{Feedback: } We adopt a power-law model to parameterize the stellar feedback, treating the disk and spheroidal components separately. The outflow rate is calculated using the following equation:

\begin{equation}
    \dot{M}_\mathrm{outflow} = \left(\frac{V_\mathrm{outflow}}{V}\right)^{\alpha_\mathrm{outflow}} \frac{{\dot{E}}}{E_\mathrm{canonical}}.
\end{equation}
Here, $V_\mathrm{outflow}$ is the characteristic velocity, set to 250 km/s and 100 km/s for the disk and spheroid components, respectively. The tunable exponent $\alpha_\mathrm{outflow}$ is set to 2 for both components. $E$ is the rate of energy input from the stellar populations and $E_\mathrm{canonical}$ is the total energy input by a canonical stellar population, normalized to $1 \mathrm{M}_\odot$ after infinite time.

\textit{Reionization model: } Our reionization model employs a methodology similar to that introduced by \cite{Benson2020}. Specifically, we assume that the intergalactic medium is instantaneously and fully reionized at redshift z = $9.97$, as determined by \cite{Hinshaw2013}. This instantaneous reionization results in a rapid photoheating of the IGM to T = $3 \times 10^{4}\,\mathrm{K}$, followed by a cooling such that the temperature at redshift z = $0$ drops to T = $1 \times 10^{3}\,\mathrm{K}$, resulting in an electron scattering optical depth of $0.0633$ in this model\footnote{We would like to note that the electron scattering optical depth utilized in this study slightly deviates from the assumptions of \citetalias{2020MNRAS.493.1614F} model, but remains within 1.3$\sigma$ of their results. The Planck 2018 \citep{2020A&A...641A...6P} results were employed in their analysis for this purpose.}.

\textit{Accretion mode: } Accretion of baryonic component into halos is computed using the filtering mass prescription of \cite{Naoz-Barkana2007}. In this prescription, it is assumed that the gas mass content of the halos is given by:
    \begin{equation}
        M_\mathrm{g}(M_\mathrm{200b},M_\mathrm{F}) = (\Omega_\mathrm{b} / \Omega_\mathrm{M}) f(M_\mathrm{200b}/M_\mathrm{F}) M_\mathrm{200b},
        \label{eq:massAccreted}
    \end{equation}
where $M_\mathrm{F}$ is the filtering mass, as first introduced by \cite{Gnedin2000} (here defined following \citealt{Naoz-Barkana2007}), $M_\mathrm{200b}$ is the halo mass defined by a density threshold of 200 times the mean background density, and $\Omega_\mathrm{b}$ and $\Omega_\mathrm{M}$ are baryon and total matter densities as a fraction of the critical density, and
    \begin{equation}
      f(x) = [1-(2^{1/3}-1) x^{-1}]^{-3}.
    \end{equation}

The accretion rate onto the halo is therefore assumed to be
    \begin{equation}
      \dot{M}_\mathrm{g} = \frac{\Omega_\mathrm{b}} {\Omega_\mathrm{M}} \frac{\mathrm{d}} {\mathrm{d} M_\mathrm{200b}} \left[
      f(M_\mathrm{200b}/M_\mathrm{F}) M_\mathrm{200b} \right] \dot{M}_\mathrm{total}.
    \end{equation}
However, in practice, three assumptions are violated. Firstly, the filtering mass is not constant in time; secondly, $M_\mathrm{total}$ does not always correspond to $M_\mathrm{200b}$; and thirdly, the growth of halos occurs through both smooth accretion and merging of smaller halos. As a result, the mass fraction in the halo will differ from $f(M_\mathrm{200b}/M_\mathrm{F})$. To address this issue, it is additionally assumed that mass flows from the hot halo reservoir to an ``unaccreted'' mass reservoir\footnote{This ``unaccreted'' reservoir represents gas in the vicinity of the halo which has been unable to accrete due to thermal pressure.} at a rate:
    \begin{equation}
    \dot{M}_\mathrm{hot} = - \frac{\alpha_\mathrm{adjust}}{\tau_\mathrm{dyn}} [M_\mathrm{hot}+M_\mathrm{unaccreted}]
    [f_\mathrm{accreted}-f(M_\mathrm{200b}/M_\mathrm{F})],
    \end{equation}
where $\alpha_\mathrm{adjust}=0.3$ is chosen to ensure that the relation between gas mass and halo mass in equation \ref{eq:massAccreted} is approximately maintained, $\tau_\mathrm{dyn}$ is the dynamical timescale, $M_\mathrm{unaccreted}$ is the mass in the unaccreted reservoir, and $f_\mathrm{accreted}$. By making these adjustments in the model, the effects of the increased gas pressure in the IGM on accretion into the CGM are accounted for.

\textit{Angular momentum: } To track the angular momentum content of halos (and their constituent gas) we adopt the random-walk model first proposed by \cite{Vitvitska2002} and developed further by \citet[][readers are encouraged to consult this paper for more detailed information]{Benson2020Behrens} which predicts the spins of dark matter halos from their merger histories. According to this model, the acquisition of angular momentum by halos occurs through the cumulative effects of subhalo accretion. By incorporating this angular momentum prescription, we can effectively reproduce the distribution of spin parameters observed in N-body simulations \citep{Benson2020Behrens}. This approach is advantageous in accounting for the intricate processes associated with halo formation and evolution (specifically for the lower mass objects), thereby providing a more accurate representation of the dynamics and properties of the simulated halos.

In \cite{Benson2020Behrens} the model was applied only to very well-resolved halos. Since, in this work, we want to explore galaxy formation in very low-mass halos---much closer to the resolution limit of the merger trees---it becomes imperative to consider the unresolved mass accretion into the halos and the corresponding alterations in angular momentum, particularly for the lower mass range. Therefore, we include an additional stochastic contribution to the angular momentum from unresolved accretion. This represents the fact that the angular momentum vector of a halo will diffuse away from zero in a random walk even if the mean angular momentum contributed by unresolved accretion is zero. The three components of the angular momentum vector of unresolved accretion are treated as independent Wiener processes with time-dependent variance that scales as the characteristic angular momentum of the halo. Specifically, each component of the angular momentum vector obeys:
\begin{equation}
 J_\mathrm{i}(t_2) = J_\mathrm{i}(t_1) + \sigma \sqrt{\Delta J^2_\mathrm{v}} N(0,1)
\end{equation}    
where $\Delta J^2_\mathrm{v}$ represents the change in (the square of) the characteristic angular momentum of the halo, $J_\mathrm{v} = M_\mathrm{v}(t) V_\mathrm{v}(t) R_\mathrm{v}(t)$, due to unresolved accretion. Here $M_\mathrm{v}(t)$, $V_\mathrm{v}(t)$, and $R_\mathrm{v}(t)$ are the virial mass, velocity, and radius respectively, $\sigma^2$ represents the variance in angular momentum per unit increase in $J_\mathrm{v}^2$, and $N(0,1)$ is a random variable distributed as a standard normal.

Making the approximation that the characteristic angular momentum scales in proportion to mass\footnote{In detail this is not correct, as there is also some dependence on the change in redshift across the timestep due to the dependence of virial densities on redshift. In practice, we ignore this dependence and absorb such effects into the parameter $\sigma$.} we can write
\begin{equation}
\Delta J^2_\mathrm{v} \approx J_\mathrm{v}^2(t_2) - J_\mathrm{v}^2(t_1) \left\{ \frac{M(t_1)+M_\mathrm{r}}{M(t_1)} \right\}^2 = J_\mathrm{v}^2(t_2) - J_\mathrm{v}^2(t_1) \left\{ \frac{M(t_2)-M_\mathrm{u}}{M(t_1)} \right\}^2
\end{equation}
where $M_\mathrm{r}$ and $M_\mathrm{u}$ are the resolved and unresolved mass accreted between times $t_1$ and $t_2$ respectively.

This model captures the idea that the increase in angular momentum from a merging event should be of order $\Delta M V_\mathrm{v}(t) R_\mathrm{v}(t)$ (since merging halos have velocities which scale with $V_\mathrm{v}(t)$ and occur at separation $R_\mathrm{v}(t)$). Additionally, because this is a Wiener process the resulting distribution of $J_\mathrm{i}(t)$ at any given time is independent of the number of steps used to get from $t=0$ to that time. (That is, the results are independent of how finely we sample the mass accretion history of each halo.) We find that $\sigma^2= 0.001$ results in reasonably good agreement between predicted and observed galaxy sizes (as discussed in more detail in \S\ref{size-mass}).

\subsection{Dark Matter halo evolution in {\sc Galacticus}}\label{app:model_A2}

Unlike the approach taken by \cite{Weerasooriya2022}, who utilized merger trees extracted from N-body simulations, this study utilizes the {\sc Galacticus} framework for the evolution of both the DM and baryonic components within the halos. Here, we provide a brief overview of this process. {\sc Galacticus} constructs merger trees for dark matter halos backwards in time using the algorithm of \cite{Cole2000}, along with the modified merger rates found by \cite{2017MNRAS.467.3454B}, which were constrained to match the progenitor mass functions in the MultiDark \citep{2016MNRAS.457.4340K} N-body simulation suite (see \href{https://github.com/galacticusorg/galacticus/wiki/Constraints:-Dark-matter-progenitor-halo-mass-functions}{here}). It then evolves the properties of the halos forward in time. When halos merge, the more massive one becomes the host, while the smaller one becomes a subhalo orbiting within it. Subhalos are initialized at the host's virial radius, positioned isotropically at random, with velocities drawn from distributions predicted by cosmological simulations. In this work, we adopt parameters from \cite{Jiang2015} and best-fit values from \cite{Benson2020Behrens}. The positions and densities of subhalos are tracked over time, accounting for dynamical friction, tidal stripping, and tidal heating until specified disruption criteria are met \citep{Pullen2014,Yang2020}. To enable rapid simulation, interactions between subhalos are ignored (see \cite{Penarrubia2005} for a justification of this approximation), and subhalos are disrupted if their bound mass falls below $10^7 \mathrm{M}_\odot$ or they pass within a distance from the host halo center equal to  1\% of the host's virial radius. For a more comprehensive explanation, refer to \cite{Yang2020}; we also refer the reader to the recent comparison between {\sc Galacticus} predictions and Symphony simulations in \cite{2023ApJ...945..159N}.

Here we explain further the nonlinear dynamical processes that govern the subhalo orbital evolution within the host halo.

\textit{Dynamical friction:} causes a subhalo to decelerate as it traverses the dark matter particles of the host halo. This is modeled using the Chandrasekhar formula \citep{Chandrasekhar1943}, assuming a Maxwell–Boltzmann distribution of host particles (see eq.~(1) in \citealt{Yang2020}). Which introduce our first free parameter, the ``Coulomb logarithm (ln $\Lambda$)''.

\textit{Tidal stripping:} removes mass from the subhalo that lies beyond the tidal radius \citep{King1962, Frank2018}, where the tidal force from the host exceeds the subhalo's self-gravity. This is modeled following \cite{Zentner2005}, with mass being removed outside the tidal radius over an orbital timescale (see eq.~(5) in \citealt{Yang2020}). Our second free parameter, $\alpha$, controls the strength of tidal stripping.

\textit{Tidal heating:} injects energy into the subhalo through rapidly varying tidal forces, causing it to expand. This is modeled using the impulse approximation with an adiabatic correction factor and a tidal tensor time integral decay term (see eq.~(8) in \citealt{Yang2020}). The exponent $\gamma$ controls the adiabatic correction term, as discussed by \cite{Gnedin1999}. The value of $\gamma$ is somewhat uncertain, with \cite{Gnedin1999} finding $\gamma$ = 2.5 (which was used by \citealt{Pullen2014}), while theoretical considerations predict $\gamma$ = 1.5 in the slow-shock regime \citep{Gnedin1999, Weinberg1994a, Weinberg1994b}. The heating coefficient, $\epsilon_{\mathrm{h}}$, which accounts for the higher order heating effects, is treated as a free parameter. This model was later improved by incorporating second-order terms in the impulse approximation for tidal heating (see eq.~(4) in \citealt{Benson2022}), allowing for an accurate match to the tidal tracks observed in high-resolution N-body simulations (refer to \cite{Benson2022} for further details).

An initial calibration of these free parameters was performed by \cite{Yang2020} using an MCMC fitting workflow to thoroughly explore the parameter space with high efficiency. For the purpose of this study, we adopt ln $\Lambda$ = 1.35, $\epsilon_{\mathrm{h}}$ = 2.70, and $\alpha$ = 2.95. We approximate these values for the choice of $\gamma$ = 1.5 (as used in the updated tidal heating model of \citealt{Benson2022}) by interpolating between the cases of $\gamma$ = 0.0 and 2.5, using the Caterpillar simulations as calibration target. 

\section{Occupation fraction -- comparison with hydrodynamical simulations}\label{app:occ}

%%%%%%%%%%%%%%%%
\begin{figure*}
    \includegraphics[width=2\columnwidth]{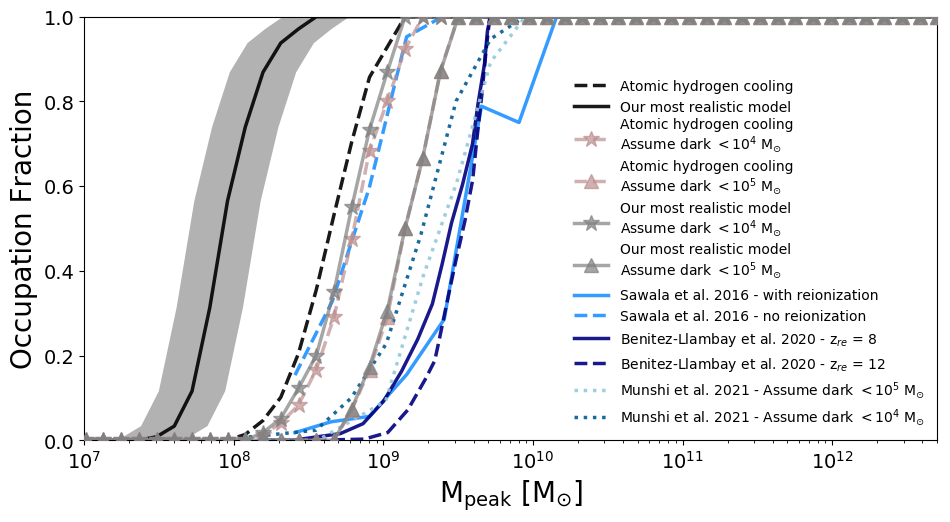}
    \caption{Comparison to occupation fractions predicted by hydrodynamical simulations. Solid black line with gray shading shows the result of our preferred model, while the long-dashed thick black curve indicates the median occupation fraction when not including $H_2$ cooling. Results from \protect \cite{Sawala2016focc} and \protect \cite{Benitez2020} are presented with blue and light-blue curves. Dotted lines with blue hues illustrate results from \protect \cite{Munshi2021}. For all hydrodynamical simulations, the resolution to call a halo ``dark" varies between $M_* < 10^4 \rm - 10^5\; \rm M_\odot$, as indicated by the labels. To imitate resolution effects from simulations, we apply further cuts to our preferred and no-$H_2$ cooling models considering as ``dark" galaxies with $M_* < 10^4\; \rm M_\odot$ (starred symbols) and  $M_* < 10^5\; \rm M_\odot$  (triangle symbols). Including resolution cuts, in particular $M_* < 10^5\; \rm M_\odot$, brings our model in closer agreement with prediction from other simulations. However, we still predict a higher occupation fraction than these models.}
    \label{fig:focc_hydro}
\end{figure*}
%%%%%%%%%%%%%%%%

Fig.~\ref{fig:focc_hydro} shows a comparison of our occupation fraction predictions to those from several hydrodynamical simulations. At face value, our model predicts the occupation of substantially lower-mass halos compared to simulations. The solid black line with gray shading indicates our preferred model, while lines in blue hues represent predictions from hydrodynamical simulations from \citet{Sawala2016focc}, \citet{Benitez2020} and \citet{Munshi2021} (see legends). However, there are two main factors that complicate this comparison. First, the physics included (and its implementation) vary substantially from model to model. We have checked in detail the different physics being implemented in {\sc Galacticus} compared to these other simulations and conclude that the inclusion of H$_2$ cooling, likely accounts for the majority of the difference between our predictions and those of some hydrodynamical simulations. We show in the dashed black line how our predictions would change if only atomic hydrogen cooling was included. As expected, it lowers the occupation fraction of low mass halos, bringing our model into closer agreement with simulations, although it still shows a somewhat higher fraction of halos with a luminous component compared to hydrodynamical simulations.

The second factor complicating the comparison between our model and simulations is numerical resolution. In simulations, the particle mass and force resolution impose a limit in the formation of ``luminous'' galaxies, which tend to occur in higher mass halos than those resolved in our semi-analytical model. For instance, \citet{Munshi2021} clearly shows that the occupation fraction, defined as the halo mass where 50\% of halos hosts a luminous component, might change by up to 1 dex in halo mass by varying the minimum $M_\star$ resolved. To examine this behavior, we impose two cuts to our preferred and no-H$_2$ cooling models: i) consider as dark all halos with a stellar mass below $M_\star<10^4\; \rm M_\odot$ (curve with triangle symbols) and, ii) $M_\star<10^5\; \rm M_\odot$ (curve with starred symbols). Interestingly, when applying these relatively ``bright'' cuts, the difference between models with and without H$_2$ cooling disappears. This highlights that the physics of molecular hydrogen cooling is only important when modeling the low mass end of the ultra-faint galaxies, or  galaxies with $M_\star<10^4\; \rm M_\odot$. It is also worth highlighting that imposing these cuts to mimic resolution effects brings our predicted occupation fractions into much closer agreement with simulations, suggesting that hydrodynamical results may be affected by the definition of the occupation fraction, which in turn is limited by resolution in these studies.

\section{Comparison to other SAMs}\label{app:otherSAMs}

Previous studies of the MW satellite galaxies in the context of the CDM cosmology have been made using other, similar SAM frameworks. Earlier studies by \cite{Benson2002, Somerville2002, Kravtsov2004} only compared with the ``classical" satellite population of dwarfs around MW (down to M$_V$ = -8.8) due to the lack of resolution. More recent studies have pushed this limit further by using N-body simulations with better resolution as benchmarks for the formation and evolution of the subhalos that are used by SAMs to host the MW and its satellite population. 

One of the N-body simulations used is the Via Lactea II simulation, which was adopted by \cite{Busha2010} to explore the effects of inhomogeneous reionization on the population of MW satellites. The availability of larger and smaller volume (lower and higher resolutions) realizations of the simulation allowed these authors to assess spatial variations in the epoch of reionization. Their galaxy evolution model was much more simplistic than that employed in this work in general, but their luminosity function predictions seem to qualitatively agree with our simplest model (i.e. the model including atomic hydrogen cooling and assuming a reionization redshift of $\sim$10). A similar N-body simulation was used by \cite{Munoz2009} who adopted a slightly more involved approach to account for star formation at different times. They found agreement with the observed satellite luminosity function (those discovered prior to SDSS and the ultra-faint sample found in the SDSS DR5) and showed that molecular hydrogen cooling is important for producing the correct abundance of low-luminosity satellites, although their molecular hydrogen cooling model is different from the one used in our study and the effect of which is stopped at $z = 20$ when they assume molecular hydrogen to be dissociated.

Another N-body simulation frequently used for the study of MW-analogs via SAMs is the Aquarius simulation. \cite{Maccio2010} compares results from three different semi-analytic models of galaxy formation (SAMs by \citealt{Kang2009, Somerville2008}, and MORGANA first presented in \citealt{Monaco2007}) applied to high-resolution N-body simulations (Aquarius). The subhalo information was \emph{not} used to determine the evolution of satellite galaxies (e.g. to determine merging timescales). To add a suitable reionization-induced suppression of galaxy formation, the \cite{Gnedin2000} filtering mass prescription is added to each model, with a reionization history taken from \cite{Kravtsov2004}. They found that all three models can achieve a reasonable match to the observed satellite luminosity function with a reionization epoch of $z = 7.5$. However, they note that the original filtering mass prescription overestimates the suppressing effects of reionization. Adopting the currently favored suppression (which becomes effective in haloes with characteristic velocities below $\sim$30 km/s), they found that a higher redshift, $z = 11$, of reionization is required to restore a good match to the data. \cite{Maccio2010} explored the roles of various physical ingredients in their models in achieving this match. In particular, they found that the inclusion of supernova feedback is crucially important---without it far too many luminous galaxies are formed. The Aquarius simulation was also used in the study of \cite{Li2010} where they apply an updated version of the ``Munich model'' described by \cite{DeLucia2007} (with updates to the reionization and feedback prescriptions) to study MW satellites. The cooling model used in this study is similar to that used in this work \citep{White-Frenk1991}. It is important to note that cooling via molecular hydrogen was not included, under the assumption that H$_2$ is efficiently photodissociated. Given this difference they are still able to reproduce the luminosity function for MW satellites, but their mass-metallicity relation does not seem to predict the plateau observed at the lower mass end. Although not directly stated in their results, we can infer from their Fig.~15 a threshold of peak halo mass above which all of their subhalos are luminous (this threshold is $\sim$10$^9$ M$_{\odot}$) which approximately agrees with our model where we include only atomic hydrogen cooling (as expected), but is clearly not able to produce the occupation fractions inferred from observation by recent studies (see results from \citealt{Nadler2020}).

The work conducted by \cite{Font2011} using the GALFORM model is closest to our approach in terms of the range of physics modelled and the detail of the treatment, e.g., inclusion of H$_2$ cooling and the evolution of the IGM is essentially that described in \cite{Benson2006}, inclusion of UV background radiation by \cite{Haardt2001} (note that all the analogous models employed in this study have undergone substantial revisions). However, they introduce a simplistic model to account for the impacts of local photoheating from local sources which appears to overestimate the contribution of local photons to the suppression of low-mass satellites (by pushing the temperature rise in the local IGM to significantly earlier epochs, leading to a more substantial suppression of gas accretion). Notably, our model does not directly encompass photoheating; rather, its effect is encapsulated through the incorporation of our reionization model and a filtering mass within our model for accretion of IGM gas into halos, thereby regulating the post-reionization temperature of the intergalactic medium. Interestingly their model foresees a distinct plateau in the mass-metallicity relation, a prediction that resonates with our model's outcomes and aligns with the current observational inferences.

Overall, {\sc Galacticus} employs the extended Press-Schechter formalism to construct merger trees, allowing it to transcend resolution limitations associated with N-body simulations (although it has the capacity to utilize merger trees derived from N-body simulations). It is important to note that for the purpose of this study we are resolving progenitor halos down to 10$^7$ M$_{\odot}$, which, as briefly discussed in Appendix~\ref{App:resolution}, gives us sufficient resolution. Additionally, the H$_2$ cooling model along with the FG20 UV background radiation introduced in section~\ref{methods} are updated versions of ones utilized in previous studies. Additionally, we have explored the effects of inclusion of an IGM metallicity model. 
It is worth noting that different SAMs, adopting various models for these key physical processes, yield comparable outcomes through minor calibrational adjustments (specifically evident in studies on luminosity function, which tend to align with observations). This might suggest the presence of degeneracies in the way in which different physical processes can affect the predictions of each model (as also suggested by \citealt{Font2011}). However, a comprehensive analysis of multiple observables rather than a singular property observed in the galaxy populations under scrutiny could potentially untangle these degeneracies. This is what we aim to accomplish in this paper by presenting a range of models and discerning differences across various observables such as the luminosity function, mass-metallicity relation, size-mass relation, and velocity dispersion-mass relation, in addition to exploring inferred theoretical properties such as the SMHM relation and occupation fraction.

\section{Resolution study}\label{App:resolution}

\begin{figure}
    \includegraphics[width=\columnwidth]{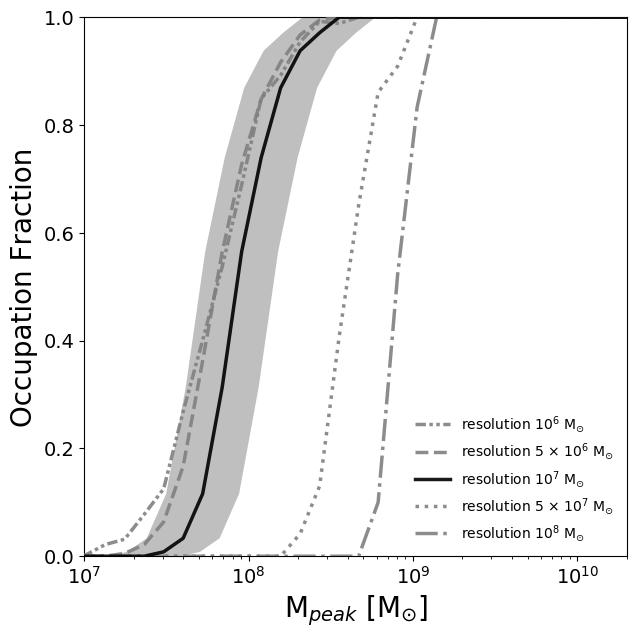}
    \caption{Impact of resolution on the predicted occupation fraction as a function of the peak halo mass. Line types depict different resolutions, with the grey double-dotted-dashed, dashed, dotted, and dashed-dotted lines corresponding to resolutions of $10^6$, $5 \times 10^6$, $5 \times 10^7$ and $10^8 \mathrm{M}_\odot$, respectively. The black solid line represents our fiducial $10^7 \mathrm{M}_\odot$ resolution, along with the uncertainty of the peak mass measurements from our results, which is depicted by the gray shaded region.}
    \label{fig:convergence_focc}
\end{figure}

In this section, we conducted tests to evaluate the performance of our model at different resolutions. We specifically assessed the impact of resolution on the prediction of the occupation fraction. Our results demonstrate that the accuracy of the occupation fraction predictions is not hindered by the resolution of $10^7 \,\mathrm{M}_{\odot}$ used in this study (illustrated by the black curve, with the corresponding dispersion indicated by the grey shaded region in Fig.~\ref{fig:convergence_focc}). To show this, we investigated higher resolutions, including $5 \times 10^6 \mathrm{M}_{\odot}$ (illustrated by the dashed grey curve) and $10^6 \mathrm{M}_{\odot}$ (depicted by the double-dotted-dashed grey curve\protect\footnote{Due to computational limitations, these results are derived from simulations with only four merger trees, sampled from the same mass range as other cases. Given the negligible predicted scatter for the halo-to-halo cases, we anticipate minimal impact on the calculated median occupation fraction.}), revealing consistent occupation fraction predictions within the statistical uncertainty of our results. However, it should be noted that our model predicts that only a fraction of our subhalos with masses below $\sim$ $2 \times 10^8 \,\mathrm{M}_{\odot}$ host a luminous component. As a result, going above a resolution of $10^7 \,\mathrm{M}_{\odot}$ (such as resolutions of $5 \times 10^7 \mathrm{M}_{\odot}$ and $10^8 \,\mathrm{M}_{\odot}$ depicted by the dotted and dashed-dotted lines on the plot) would significantly impact the results.

%%%%%%%%%%%%%%%%
%%%%%%%%%%%%%%%%

\begin{figure}
    \includegraphics[width=\columnwidth]{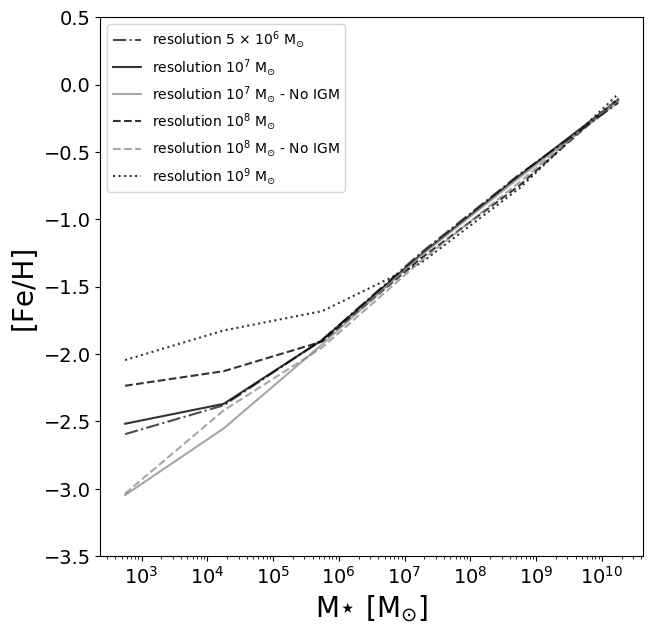}
    \caption{Effect of resolution on the mass -- metallicity relation. The mass-metallicity relation is shown with various black line styles representing different resolutions. Resolution $10^7 \mathrm{M}_\odot$ is represented by the solid black line, while resolutions $5 \times 10^6$, $10^8$ and $10^9 \mathrm{M}_\odot$ are depicted by the dotted-dashed, dashed and dotted black lines, respectively. Additionally, the corresponding models without the inclusion of IGM metallicity are shown with gray lines, with solid and dotted lines representing resolutions $10^7$ and $10^8 \mathrm{M}_\odot$, respectively.}
    \label{fig:convergence_metal}
\end{figure}

We also examined the influence of resolution on the predicted metallicity of the satellites. As depicted in Fig.~\ref{fig:convergence_metal}, at higher stellar masses for the subhalos, we did not observe substantial differences resulting from resolution changes. However, at the lower mass range, altering the resolution introduced some variations in the metallicity predictions. These discrepancies could be attributed to the effects of bias in selecting the low stellar mass population due to the sharp resolution-induced cutoff in the SMHM relation within our results. In such cases, lower resolution would lead to a subhalo population with biased higher stellar masses (due to the resolution cutoff), which can statistically shift the median metallicity towards larger values. Additionally, we excluded the IGM metallicity from our model and compared its impact. The results demonstrated a similar behavior with slightly reduced significance.

%%%%%%%%%%%%%%%%
%%%%%%%%%%%%%%%%

%%%%%%%%%%%%%%%%%%%%%%%%%%%%%%%%%%%%%%%%%%%%%%%%%%

% Don't change these lines
\bsp	% typesetting comment
\label{lastpage}
\end{document}